\documentclass[twocolumn]{article}
\pdfoutput=1

\usepackage[T1]{fontenc}
\usepackage[utf8]{inputenc}
\usepackage[usenames,dvipsnames]{color}
\usepackage{graphicx}
\usepackage[cmex10]{amsmath}
\usepackage{hyperref}
\usepackage{xspace}
\usepackage[caption=false,font=footnotesize]{subfig}
\usepackage{amssymb}
\usepackage{todonotes}
\usepackage[vlined,linesnumbered]{algorithm2e}
\usepackage{booktabs}
\usepackage{supertabular}
\usepackage[margin=1.8cm]{geometry}

\newcommand{\ie}{i.\,e.\xspace}
\newcommand{\eg}{e.\,g.\xspace}

\definecolor{bg}{rgb}{0.985,0.985,0.985}

\hypersetup{breaklinks=true,
            pdfauthor={Michael Hamann, Gerd Lindner, Henning Meyerhenke, Christian L. Staudt, Dorothea Wagner},
            pdftitle={Structure-Preserving Sparsification Methods for Social Networks}
            colorlinks=true,
            citecolor=PineGreen,
            urlcolor=MidnightBlue,
            linkcolor=PineGreen,
            pdfborder={0 0 0}}
\definecolor{bg}{rgb}{0.985,0.985,0.985}

\title{Structure-Preserving Sparsification Methods \\ for Social Networks\thanks{Parts of this paper have been published in preliminary form in~\cite{DBLP:conf/asunam/LindnerSHMW15}.}}
\author{Michael~Hamann \and Gerd~Lindner \and Henning~Meyerhenke \and Christian~L.~Staudt \and Dorothea~Wagner}
\date{}                                           

\begin{document}
\maketitle

\begin{abstract} \small\baselineskip=9pt 
Sparsification reduces the size of networks while preserving structural and statistical properties of interest.
Various sparsifying algorithms have been proposed in different contexts.
We contribute the first systematic conceptual and experimental comparison of \textit{edge sparsification} methods on a diverse set of network properties.
It is shown that they can be understood as methods for rating edges by importance and then filtering globally or locally by these scores.
We show that applying a local filtering technique improves the preservation of all kinds of properties.
In addition, we propose a new sparsification method (\textit{Local Degree}) which preserves edges leading to local hub nodes.
All methods are evaluated on a set of social networks from Facebook, Google+, Twitter and LiveJournal with respect to network properties including diameter, connected components, community structure, multiple node centrality measures and the behavior of epidemic simulations.
In order to assess the preservation of the community structure, we also include experiments on synthetically generated networks with ground truth communities.
Experiments with our implementations of the sparsification methods (included in the open-source network analysis tool suite NetworKit)
show that many network properties can be preserved down to about 20\% of the original set of edges for sparse graphs with a reasonable density.
The experimental results allow us to differentiate the behavior of different methods and show which method is suitable with respect to which property.
While our Local Degree method is best for preserving connectivity and short distances, other newly introduced local variants are best for preserving the community structure.

~\\[0.5ex]
\end{abstract}

\section{Introduction}

\subsection{Context}

Complex networks have nontrivial structures and statistical properties and are often represented by graphs.
Such data models have been employed in countless domains based on the observation that the structure of relationships yields insights into the composition and behavior of complex systems~\cite{costa2011analyzing}.
Many concepts were pioneered in the study of social networks, in which edges represent social ties between social actors.
Most real-world complex networks, including social networks, are already sparse in the sense that for $n$ nodes the edge count $m$ is asymptotically in $O(n)$.
Nonetheless, typical densities lead to a computationally challenging number of edges.
Here we pursue the goal of further sparsifying such networks by retaining just a fraction of edges (sometimes called a ``backbone'' of the network), while showing experimentally that important properties of networks can be preserved in the process.

Potential applications of network sparsification are numerous.
One of them is information visualization: 
Even moderately sized networks turn into ``hairballs'' when drawn with standard techniques, as the amount of edges is visually overwhelming.
In contrast, showing only a fraction of edges can reveal network structures to the human eye if these edges are selected appropriately. 
Sparsification can also be applied as an acceleration technique: By disregarding a large fraction of edges that are unimportant for the task, running times of graph and network analysis algorithms can be reduced.
Many other possible applications arise if we think of sparsification as lossy compression.
Large networks can be strongly reduced in size if we are only interested in certain structural aspects that are preserved by the sparsification method.
From a network science perspective, sparsification can yield valuable insights into the importance of relationships and the participating nodes: 
Given that a sparsification method tends to preserve a certain property, the method can be used to rank or classify edges, discriminating between essential and redundant edges.

The core idea of the research presented here is that not all edges are equally important with respect to properties of a network: For example, a relatively small fraction of long-range edges typically act as shortcuts and are responsible for the small-world phenomenon in complex networks. 
The importance of edges can be quantified, leading to \textit{edge scores}, often also referred to as \textit{edge centrality values}.
In general, we subsume under these terms any measure that quantifies the importance of an edge depending on its position within the network structure.
Sparsification can then be broken down into the stages of (i) edge scoring and (ii) filtering the edges using a global score threshold.

Despite the similar terminology, our work is only weakly related to a line of research in theoretical computer science where \textit{graph sparsification} is understood as the reduction of a dense graph ($\Theta(n^2)$ edges) to a sparse ($O(n)$ 
edges) or nearly-sparse graph  while provably preserving properties such as spectral properties (\eg~\cite{batson2013spectral}).
The networks of our interest are already sparse in this sense.
With the goal of reducing network data size while keeping important properties, our research is related to a body of work that considers sampling from networks (on which~\cite{ahmed2014network} provides an extensive overview).
Sampling is concerned with the design of algorithms that select edges and/or nodes from a network.
Here, node and edge sampling methods must be distinguished:
For node sampling, nodes \emph{and} edges from the original network are discarded, while edge sampling preserves all nodes and reduces the number of edges only.
The literature on node sampling is extensive, while pure edge sampling and filtering techniques have not been considered as often.
A seminal paper~\cite{Leskovec2006} concludes that node sampling techniques are preferable, but considers few edge sampling techniques.
The study presented in~\cite{ebbes2008sampling} looks at how well a sample of 5\%-20\% of the original network preserves certain properties, and is mainly focused on node sampling through graph exploration. It concludes that random walk-based node sampling works best on complex networks, but does so on the basis of experiments on synthetic graphs only and compares only with very simple edge sampling methods. 

Only edge sampling techniques are directly comparable to our edge scoring and filtering methods.
In this work, we restrict ourselves to reducing the edge set, while keeping all nodes of the original graph.
Preserving the nodes allows us to infer properties of each node of the original graph.
This is important because in network analysis, the unit of analysis is often the individual node, \eg when a score for each user in an online social network scenario shall be computed.
With respect to the goal of accelerating the analysis, many relevant graph algorithms scale with $m$ 
so reducing $m$ is more relevant.

Another related approach is the \textit{Multiscale Backbone}~\cite{Serrano09}, which is applicable on weighted graphs only and is therefore not included in our study.
Instead of applying a global edge weight cutoff for edge filtering, which hides important structures at different scales, this approach aims at preserving them at all scales.

\subsection{Contribution}

We contribute the first systematic conceptual and experimental comparison of existing and novel edge scoring and filtering methods on a diverse set of network properties.
Descriptions and literature references for the related methods which we reimplemented are given in Section~\ref{sec:methods}, for some of them we include descriptions of how we parallelized them.
In Section~\ref{sec:methods} we also introduce our Local Degree sparsification method and Edge Forest Fire, an adaption of the existing node sparsification technique to edges.
Furthermore, we propose a local filtering step that has been introduced by \cite{Satuluri2011} for one specific sparsification technique as a generally applicable and beneficial post-processing step for preserving the connectivity of the network and most properties we consider.

Our results illuminate which methods are suitable with respect to which properties of a network.
Additionally, we take a look at emergent properties by simulating epidemic spreading on sparsified networks in comparison with the original network.
We show that our Local Degree method is best for preserving connectivity and short distances which results in a good preservation of the diameter of the network, some centrality measures and the behavior of epidemic spreading.
Depending on the network, our Local Degree method can also preserve clustering coefficients.
Considering the preservation of the community structure we show that some of the newly introduced variants with local filtering are best for preserving the community structure while the variants without local filtering do not preserve the community structure in our experiments.

Furthermore, we have published efficient parallelized implementations and a framework for such methods as part of
the \textit{\href{http://networkit.iti.kit.edu}{NetworKit}} open-source tool suite~\cite{DBLP:journals/corr/StaudtSM14}.
While our study covers various approaches from the literature, it is by no means exhaustive due to the vast amount of potential sparsification techniques. 
With future methods in mind, we hope to contribute a framework for their implementation and evaluation.

%
\section{Network Properties}
\label{sec:prop}

The structure of a complex network is usually characterized in terms of certain key figures and statistics~\cite{newman2010networks}. Decomposition of the network into cohesive regions is a frequent analysis task: 
All nodes in a \textit{connected component} are reachable from each other. A typical pattern in real-world complex networks is the emergence of a giant connected component which contains the bigger part of all nodes and is usually accompanied by a large number of very small components. \textit{Communities} are subsets of nodes that are internally dense and externally sparsely connected. The \textit{diameter} of a graph is the maximum length of a shortest path between any two nodes \cite{newman2010networks}. The observation that the diameter of social networks is often surprisingly small is referred to as the \textit{small world phenomenon}. In case of disconnected graphs, we consider the diameter of the largest component. \textit{Node centrality measures} quantify the relative importance of a node within the network structure. We consider any function which assigns to each node an attribute value of at least ordinal scale of measurement to be a node centrality measure. The distribution of \textit{degrees}, the number of connections per node,  can be seen as the simplest measure that falls under this definition. It plays  an important role in characterizing a network: Empirically observed complex networks tend to show a heavy tailed \textit{degree distribution} which follows a power-law with a characteristic exponent: $p(k) \sim k^{-\gamma}$. Such networks have been categorized as \textit{scale-free} \cite{ba-esrn-99}, referring to the fact that it is not possible to pick a node of typical degree. \textit{Clustering coefficients} are key figures for the amount of transitivity in networks. The \textit{local clustering coefficient} expresses how many of the possible connections between neighbors of a node exist, which can be treated as a node centrality measure according to the definition above \cite{newman2010networks}. \textit{Betweenness centrality} expresses the concept that a node is important if it lies on many shortest paths between nodes in the network. \textit{PageRank} \cite{Page1999} assigns relative importance to nodes according to their connections, incorporating the idea that edges leading to high-scoring nodes contribute more. While this collection is not and cannot be exhaustive, we choose these common measures for our experimental study (Section~\ref{sec:exp}).

\section{Edge Sparsification}
\label{sec:methods}

All edge sparsification methods we consider can be split up into two stages:
(i) the calculation of a score for each of the $m$ edges in the input graph (where the score is high if the edge is important) and (ii) subsequent filtering by applying a global threshold such that only edges whose score is above this threshold are kept.
In this section, we introduce Local Filtering as an optionally applicable step between the two aforementioned steps, and present the existing and new sparsification approaches we consider. For each of these methods, we show how it can be transformed into an edge score that can be used for filtering.
We also describe the used algorithms for computing these edge scores as we have developed novel parallel versions for some of them.

\subsection{Local Filtering}

\begin{figure*}[htbp]
\centering

\subfloat[Quadrilateral Simmelian Backbone]{
\includegraphics[width=0.3\textheight]{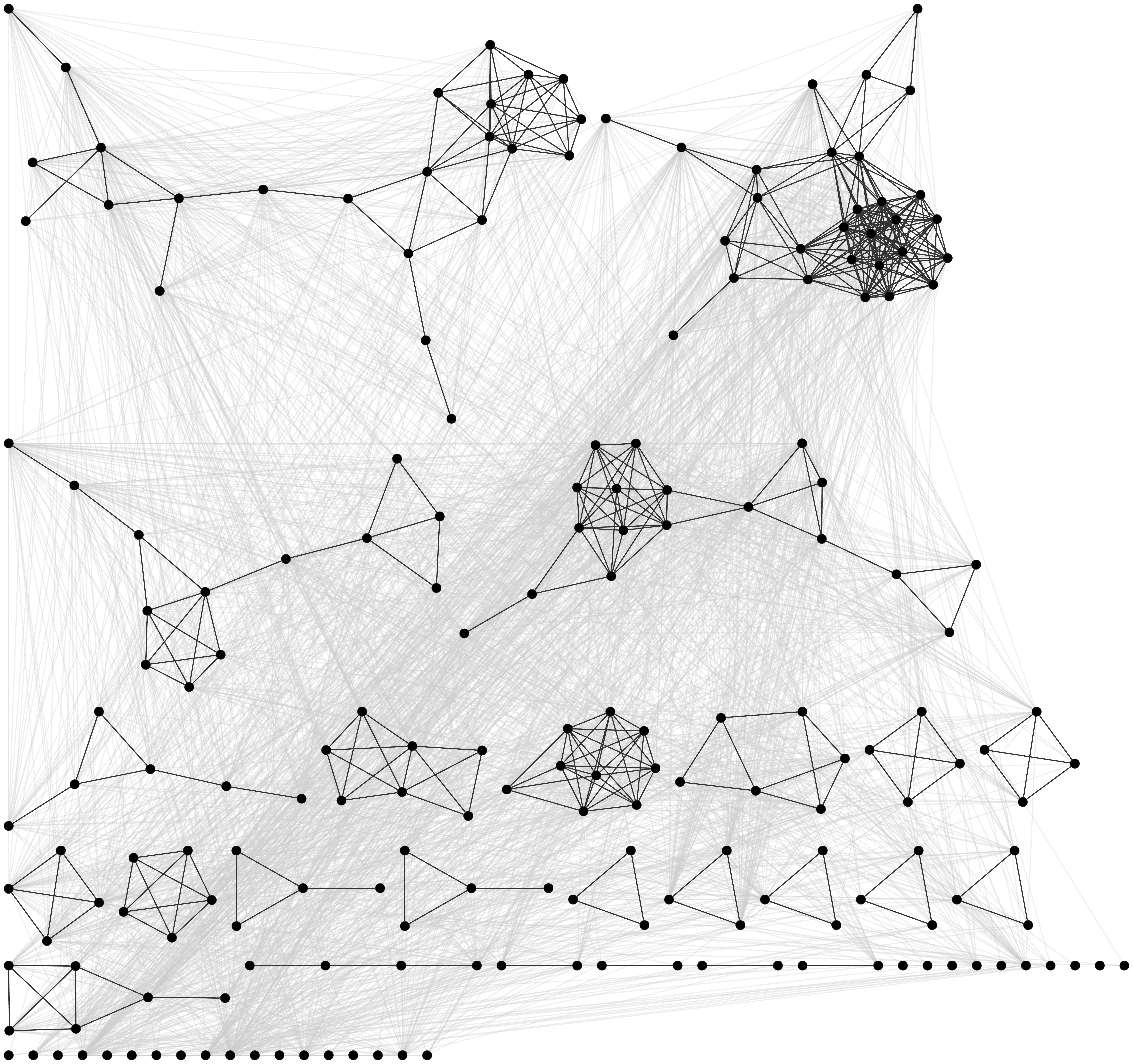}
\label{fig:jazz_qls}
}
\qquad
\subfloat[Quadrilateral Simmelian Backbone with UMST]{
\includegraphics[height=0.3\textheight]{figures/graphs/jazz-15-QLS-UMST-visone}
\label{fig:jazz_qls-umst}
}
\qquad
\subfloat[Quadrilateral Simmelian Backbone with local filtering]{
\includegraphics[height=0.3\textheight]{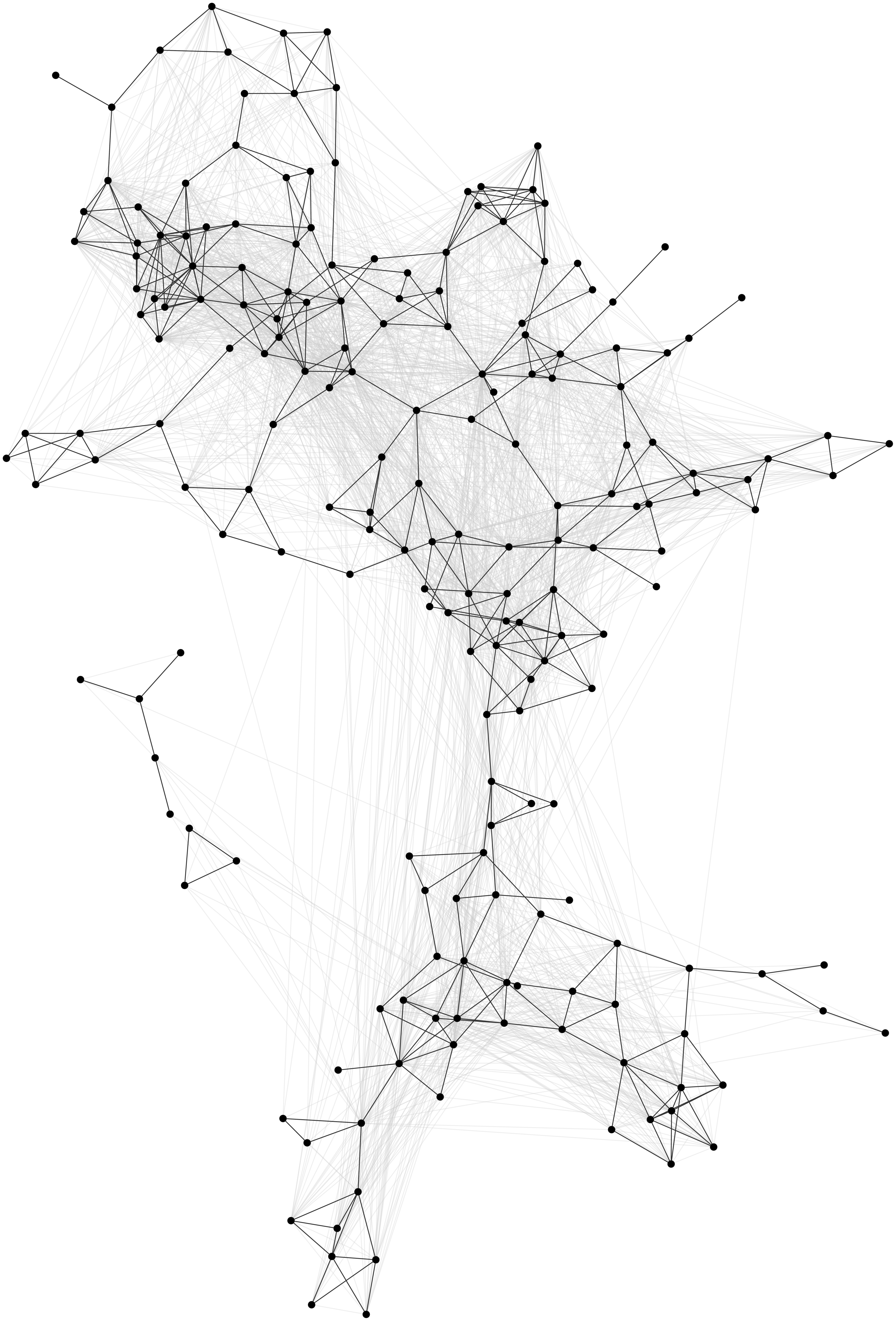}
\label{fig:jazz_lqls}
}
\caption{Drawing of the Jazz musicians collaboration network according to a variant of the Quadrilateral Simmelian Backbone with 15\% of the edges (in black)}
\end{figure*}

A problem with global filtering as described above is that methods that are based on local measures like the number of quadrangles an edge is part of tend to assign different scores in different parts of the network as \eg some parts of the network are much denser than other parts.
Sparsification techniques like (Quadrilateral) Simmelian Backbones~\cite{nocaj2014hairballs} use different kinds of normalizations of quadrangles (see below for details) in order to compensate for such differences.
Unfortunately, these normalizations still do not fully compensate for these differences.
In Fig.~\ref{fig:jazz_qls} we visualize the Jazz network \cite{GleiserJazz} with 15\% kept edges as an example.
As one can see in the figure, many nodes are isolated or split into small components, the original structure of the network (shown with gray edges) is not preserved.

Simmelian Backbones have been introduced for visualizing networks that are otherwise hard to layout.
For layouts it is important to keep the connectivity of the network as otherwise nodes cannot be positioned relative to their neighbors.
In order to preserve the connectivity, \cite{nocaj2014hairballs} keep the union of all maximum spanning trees (UMST) in addition to the original edges.
In Fig.~\ref{fig:jazz_qls-umst} we show the result when we keep the UMST.
While the network is obviously connected, much of the local structure is lost in the areas between the dense parts -- which is not surprising as we only added the union of some trees.

\cite{Satuluri2011} face a similar problem as with their sparsification technique based on Jaccard Similarity (see below for details) they want to preserve the community structure.
They propose a different solution: Each node $u$ keeps the top $\lfloor d(u)^\alpha \rfloor$ edges incident to $u$, ranked according to their similarity ($\alpha \in [0,1]$, $d(u)$ denotes the degree of $u$).
This procedure ensures that at least one incident edge of each node is retained and thus prevents completely isolated nodes.
This is equivalent to assigning each edge the score $1- \alpha$ for the minimum value of $\alpha$ such that the edge is kept in the sparsified graph and filtering globally by this new edge score.
When ranking edges, we assign a group of edges with equal score the lowest rank of the edges in the group and thus assign all equal edges the highest score.

In Fig.~\ref{fig:jazz_lqls} we show the Jazz network, sparsified again to 15\% of the edges with the Quadrilateral Simmelian Backbone method using local filtering.
With the local filtering step, the network is almost fully connected and local structures are maintained, too.
Additionally, as already Satuluri et al.\ observed when they applied local filtering to their Jaccard Similarity, the edges are much more distributed among the different parts of the network.
This means that we can still see the local structure of the network in many parts of the network and do not only maintain very dense but disconnected parts.
In our evaluation we confirm that many properties of the considered networks are indeed better preserved when the local filtering step is added.
Furthermore, we show that the local filtering step leads to an almost perfect preservation of the connected components on all considered networks even though this is not inherent in the method.
This suggests that local filtering is superior to preserving a UMST as not only connectivity but also local structures are preserved.
We therefore propose to apply this local filtering step to all sparsification methods where local filtering has not been considered yet.
In our evaluation we do not further consider the alternative of preserving a UMST as preliminary experiments have shown that adding a UMST has no significant advantage over the local filtering step in terms of the preservation of network properties.
With local filtering, our sparsification pipeline consists of the following stages: (i) calculation of an edge score, (ii) conversion of the edge score into a local edge score and (iii) global filtering.
In the evaluation we prefix the abbreviations of the local variants with ``L''.

\subsection{Sparsification Methods}
\paragraph*{Random Edge (RE). \,}

When studying different sparsification algorithms, the performance of random edge selection is an important baseline. 
As we shall see, it also performs surprisingly well.
The method selects edges uniformly at random from the original set such that the desired sparsification ratio is obtained. 
This is equivalent to scoring edges with values chosen uniformly at random.
Naturally this needs time linear in the number of edges and can be trivially parallelized.

\paragraph*{Triangles. \,}
Especially in social networks, triangles play an important role because the presence of a triangle indicates a certain quality of the relationship between the three involved nodes.
The sociological theory of Simmel~\cite{simmel1950sociology} states that ``triads (sets of three actors) are fundamentally different from dyads (sets of two actors) by way of introducing mediating effects.''
In a friendship network, it is likely for two actors with a high number of common friends to be friends as well.
Filtering globally by triangle counts tends to destroy local structures, but several of the following sparsification methods are based on the triangles edge score $T(u, v)$ that denotes for an edge $\{u, v\}$ the number of triangles it belongs to.
The time needed for counting the number of all triangles is $O(m\cdot a)$~\cite{ortmann2014triangle}, where $a$ is the graph's arboricity~\cite{Chi85}.

We use a parallelized variant of the algorithm introduced by \cite{ortmann2014triangle}.
This variant is different from the parallel variant introduced in \cite{shun2015multicore} as they need additional overhead in the form of sorting operations or atomic operations for storing local counters which we avoid.
Algorithm~\ref{alg:triangles} contains the pseudo-code for our algorithm.
The algorithm needs a node ordering.
While a smallest-first ordering that is obtained by iteratively removing nodes of minimum degree can guarantee the theoretical running time, simply ordering the nodes by degree is actually faster in practice as noticed by \cite{ortmann2014triangle}.
Therefore we use such a simple degree ordering.
While $N(u)$ denotes all neighbors of $u$, $N^+(u)$ denotes the neighbors of the node that are higher in the ordering.
Note that when using a smallest-first ordering $|N^+(u)|$ is bounded by $a$.
In contrast to \cite{ortmann2014triangle} we count each triangle three times which does not increase the asymptotic running time.
In each iteration step of the outer loop we encounter each triangle $u$ and the edges incident to $u$ are part of exactly once.
Therefore it is enough to count the triangle for the edges that are incident to $u$ and where $u$ has the higher id.
This avoids multiple accesses to the same edge by several threads, we therefore do not need any locks or atomic operations.
In the same way we could also update triangle counters per node \eg for computing clustering coefficients without additional work and without using locks or atomic operations.
Note that node markers are thread-local.

\begin{algorithm}
\ForEach(in parallel){$u \in V$}{
Mark all $v \in N(u)$\;
\ForEach{$v \in N(u)$}{
\ForEach{$w \in N^+(u)$}{
\If{$w$ is marked}{
Count triangle $u$, $v$, $w$\;
}
}
}
Un-mark all $v \in N(u)$\;
}
\caption{Parallel triangle counting}
\label{alg:triangles}
\end{algorithm}

\paragraph*{(Local) Jaccard Similarity (JS, LJS). \,}
\label{par:LocalSim}

One line of research attempts to sparsify graphs with the goal of speeding up data mining algorithms.
\cite{Satuluri2011} propose a local graph sparsification method with the intention of speedup and quality improvement of community detection.
They suggest reducing the edge set to 10-20\% of the original graph and use the Jaccard measure to quantify the overlap between node neighborhoods $N(u)$, $N(v)$ and thereby the (Jaccard) similarity of two given nodes:

\begin{equation*}
	\operatorname{JS}(u,v) = \frac{|N(u) \cap N(v)|}{|N(u) \cup N(v)|} = \frac{T(u,v)}{d(u) + d(v) - T(u, v)},
\end{equation*}

\noindent where $d(u)$ denotes the degree of $u$.
The time needed for calculating the Jaccard Similarity is the time for counting all triangles.
The authors also propose a fast approximation which runs in time $O(m)$.

For this Jaccard Similarity, \cite{Satuluri2011} propose the local filtering technique that we have already explained above and that we denote by LJS.
The time needed for calculating this local edge score is the time for calculating the Jaccard Similarity and for sorting the neighbors of all nodes, which can be done in $O(m \log(d_{\max}))$.
We process the nodes in parallel for sorting the neighbors.

This sparsification technique has also been adapted for accelerating \textit{collective classification}, \ie the task of inferring the labels of all nodes in a graph given a subset of labeled nodes~\cite{saha2013sparsification}.

\paragraph*{Simmelian Backbones (TS, QLS). \,}

The \textit{Simmelian Backbones} introduced by \cite{Nick13} aim at discriminating between edges that are placed within dense subgraphs and those between them.
The original goal of these methods was to produce readable layouts of networks.
To achieve a ``local assessment of the level of actor neighborhoods''~\cite{Nick13}, 
the authors propose the following approach, which we adapt to our concept of edge scores. Given an edge scoring method $S$ and a node $u$, they introduce the notion of a rank-ordered neighborhood as the list of adjacent neighbors sorted by $S(u,\cdot)$ in descending order.
The original \textit{(Triadic) Simmelian Backbone} uses triangle counts $T$ for $S$.
The newer \textit{Quadrilateral Simmelian Backbone} by \cite{nocaj2014hairballs} uses \textit{quadrilateral edge embeddedness}, which they define as

\begin{equation*}
Q(u,v) = \frac{q(u,v)}{\sqrt{q(u) \cdot q(v)}}
\end{equation*}
with $q(u,v)$ being the number of quadrangles containing edge $\{u,v\}$ and $q(u)$ being the sum of $q(u,v)$ over all neighbors $v$ of $u$.
They argue that this modified version performs even better at discriminating edges within and between dense subgraphs.

On top of the rank-ordered neighborhood graph that is induced by the ranked neighborhoods of all nodes, Nick et al.\ introduce two filtering techniques, a parametric one and a non-parametric one. Like Nocaj et al.\ we use only the non-parametric variant. By \textit{TS}, we denote the Triadic Simmelian Backbone and by \textit{QLS} the Quadrilateral Simmelian Backbone.
The non-parametric variant uses the Jaccard measure similar to Local Similarity but, instead of considering the whole neighborhood, they use the maximum of the Jaccard measure of the top-$k$ neighborhoods for all possible values of $k$.
While the time needed for quadrangle counting is equal to the time for triangle counting~\cite{Chi85}, the overlap and Jaccard measure calculation of prefixes needs time $O(m \cdot d_{\max})$ as it needs to be separately calculated for all edges.
We use a relatively simple implementation of the original algorithm for quadrangle counting. All neighborhoods are sorted in parallel which takes $O(m \cdot \log(d_{\max}))$ time. By using binary vectors for marking the unmatched neighbors of both incident nodes we get $O(\sum_{\{u,v\} \in E} d(u) + d(v)) = O(m \cdot d_{\max})$ for the Jaccard measure calculations which dominates the running time. We execute this calculation in parallel for all edges.

\paragraph*{Edge Forest Fire (EFF). \,}

The original Forest Fire node sampling algorithm~\cite{Leskovec2006} is based on the idea that nodes are ``burned'' during a fire that starts at a random node and may spread to the neighbors of a burning node. 
Note that contrary to random walks the fire can spread to more than one neighbor but already burned neighbors cannot be burned again. The basic intuition is that nodes and edges that get visited more frequently than others during these walks are more important.
In order to filter edges instead of nodes, we introduce a variant of the algorithm in which we use the frequency of visits of each edge as a proxy for its relevance.

Algorithm~\ref{alg:eff} shows the details of the algorithm we use to compute the edge score. The fire starts at a random node which is added to a queue.
The fire always continues at the next extracted node $v$ from the queue and spreads to neighboring unburned nodes until either all neighbors have been burned or a random probability we draw is above a given burning probability threshold $p$.
The number of burned neighbors thus follows a geometric distribution with mean $p/(1-p)$.
We use a very simple parallelization for our Edge Forest Fire algorithm.
We burn several fires in parallel with separate burn markers per thread and atomic updates of the burn frequency.
In order to avoid too frequent updates we update the global counter for the number of edges burned only after a fire has stopped burning before we start the next fire.
As the total length of all walks is hard to estimate in advance, we cannot give a tight bound for the running time.

\begin{algorithm}
\KwIn{\texttt{targetBurnRatio} $\in \mathbb{R}$, $p \in [0, 1)$}
\texttt{edgesBurnt} $\gets 0$\;
\While{\texttt{edgesBurnt} $< m \cdot $ \texttt{targetBurnRatio}}{
Add random node to queue\;
\While{queue not empty}{
$v \gets$ node from queue\;
\While{true}{
$q \gets \text{random element from} [0,1)$\;
\If{$q > p$ or $v$ has no un-burnt neighbors}{
break\;
}

$x \gets $ random un-burnt neighbor of $v$\;
Mark $x$ as burnt\;
Add $x$ to queue\;
Increase \texttt{edgesBurnt}\;
Increase burn counter of $\{v, x\}$\;
}
}
}
\caption{Edge Forest Fire}
\label{alg:eff}
\end{algorithm}

\paragraph*{Algebraic Distance (AD). \,}

Algebraic distance~\cite{chen2011algebraic} ($\alpha$) is a method for quantifying the structural distance of two nodes $u$ and $v$ in graphs. Its essential property is that $\alpha(u,v)$ decreases with the number of paths connecting $u$ and $v$ as well as with decreasing lengths of those paths. Algebraic distance therefore measures the distance of nodes by taking into account more possible paths than \eg shortest-path distance and with wider in scope than \eg the Jaccard coefficient of two nodes' immediate neighborhood. Nodes that are connected by many short paths have a low algebraic distance. It follows that nodes within the same dense subgraph of the network are close in terms of $\alpha$. Algebraic distance can be described in terms of random walks on graphs and, roughly speaking, $\alpha(u,v)$ is low if a random walk starting at $u$ has a high probability of reaching $v$ after few steps.
In a straightforward way, algebraic distance can be used to quantify the ``range'' of edges, with short-range edges (low $\alpha(u,v)$ for an edge $\{u,v\}$) connecting nodes within the same dense subgraph, and long-range edges (high $\alpha(u,v)$ for an edge $\{u,v\}$) forming bridges between separate regions of the graph. Hence, $\alpha$ restricted to the set of connected node pairs is an edge score in our terms, and can be used to filter out long- or short-range edges.
We use $1 - \alpha(u,v)$ as edge score in order to treat short-range edges as important.

$\alpha$ is computed by performing iterative local updates on $d$-dimensional ``coordinates'' of a node.
The coordinates are initialized with random values. Then, in each iteration, the coordinates are set to some weighted average of 
the old coordinates and the average of the old coordinates of all neighbors.
These updates of the node coordinates are parallelized in our code.
The algebraic distance is then any distance between the two coordinate vectors, we choose the $\ell_2$-norm.
As described in~\cite{chen2011algebraic}, $d$ can be set to a small constant (\eg 10) and the distances stabilize after tens of iterations of $O(m)$ running time each.
We choose 20 systems and 20 iterations.

\begin{figure}[h!]
\centering
\includegraphics[width=0.21\textwidth]{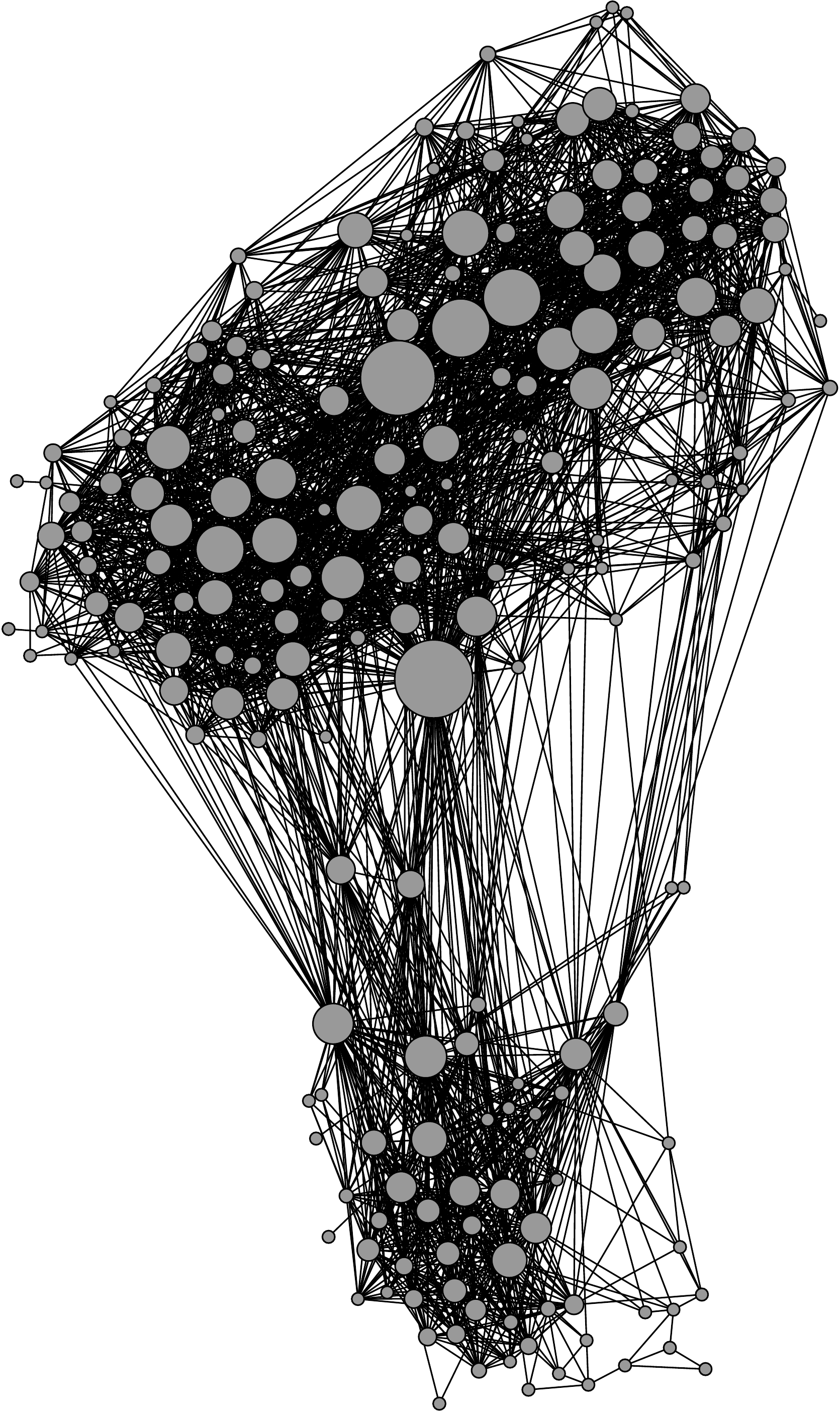}
~
\includegraphics[width=0.21\textwidth]{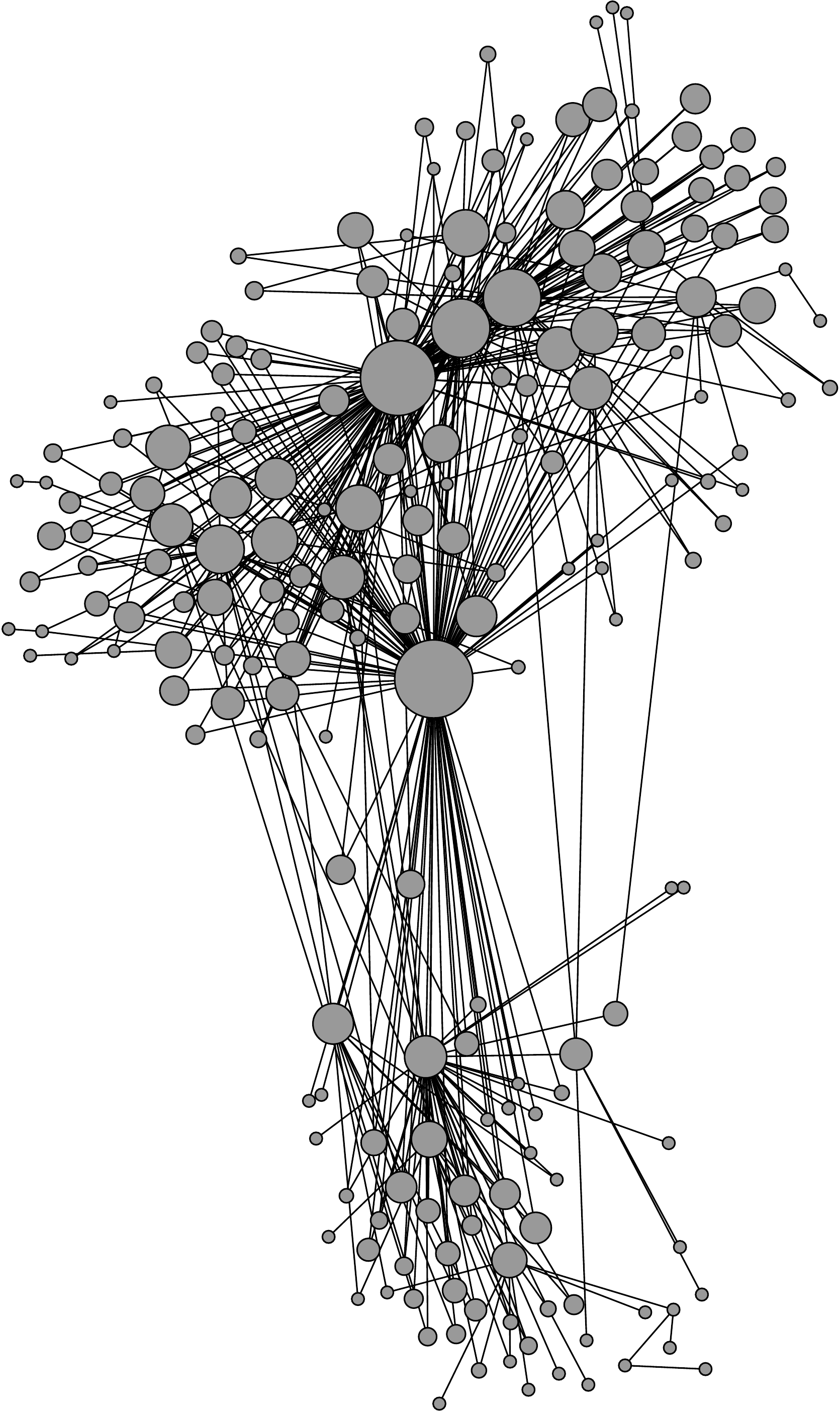}
\caption{Drawing of the Jazz musicians collaboration network and the \textit{Local Degree} sparsified version containing 15\% of edges. Node size proportional to degree.}
\label{fig:jazz}
\end{figure}

\paragraph*{Local Degree (LD). \,}
Inspired by the notion of hub nodes, \ie nodes with locally relatively high degree, and that of local sparsification, we propose the following new sparsification method:
For each node $v \in V$, we include the edges to the top $\lfloor \deg(v)^\alpha \rfloor$ neighbors, sorted by degree in descending order.
Similar to the local filtering step we explained above we use again $1-\alpha$ for the minimum parameter $\alpha$ such that an edge is still contained in the sparsified graph as edge score.
The goal of this approach is to keep those edges in the sparsified graph that lead to nodes with high degree, i.e.\ the hubs that are crucial for a complex network's topology.
The edges left after filtering form what can be considered a ``hub backbone'' of the network.
In Fig.~\ref{fig:jazz} we visualize the Jazz network \cite{GleiserJazz} as an example.

As only the neighbors of each node need to be sorted, this can be done in $O(m\log(d_{\max}))$. 
Using linear-time sorting it is even possible in $O(m)$ time.
We have decided against the linear-time variant and instead apply the sorting in parallel on all nodes.

\section{Implementation}
\label{sec:impl}

For this study, we have created efficient C++ implementations of all considered sparsification methods,
and have accelerated them using OpenMP parallelization.
All methods (with exception of the inherently sequential quadrangle counting algorithms~\cite{Chi85}) have been parallelized.
We have implemented the algorithms in \emph{NetworKit}~\cite{DBLP:journals/corr/StaudtSM14}, an interactive tool suite for scalable network analysis.
It provides a large set of graph algorithm implementations we have used for our experiments.
NetworKit combines kernels written in C++ with an interactive Python shell to achieve both high performance and interactivity, a concept we use for our implementations as well.
For community detection, we use an efficient implementation of the Louvain method with refinement that is also part of NetworKit~\cite{staudt2015engineering} as it is fast enough for the vast amount of networks that we get due to the different sparsification methods and ratios of kept edges while still detecting communities of a reasonable quality.

\textit{Gephi}~\cite{BastianHJ09Gephi} is a graph visualization tool which we use not only for visualization purposes but also for interactive exploration of sparsified graphs. To achieve said interactivity, we implemented a client for the \emph{Gephi Streaming Plugin} in NetworKit. It is designed to stream graph objects from and to Gephi utilizing the JSON format. Using our implementation in NetworKit, a few lines of Python code suffice to sparsify a graph, calculate various network properties, and export it to Gephi for drawing.
The approach of separating sparsification into edge score calculation and filtering allows for a high level of interactivity by exporting edge scores from NetworKit to Gephi and dynamic filtering within Gephi. 

For the drawings of the Simmelian Backbones we use \textit{visone}\footnote{\url{http://visone.info}}.

\section{Experimental Study}
\label{sec:exp}

Our experimental study consists of two parts.
In the first part (Sec.~\ref{sub:correlation}) we compare correlations between the calculated edge scores on a set of networks.
In the second part (Sec.~\ref{sec:quantifying}) we compare how similar the sparsified networks are to the original network by comparing certain properties of the networks. 

\subsection{Setup}
\begin{figure}
\centering
\includegraphics[width=0.9\columnwidth]{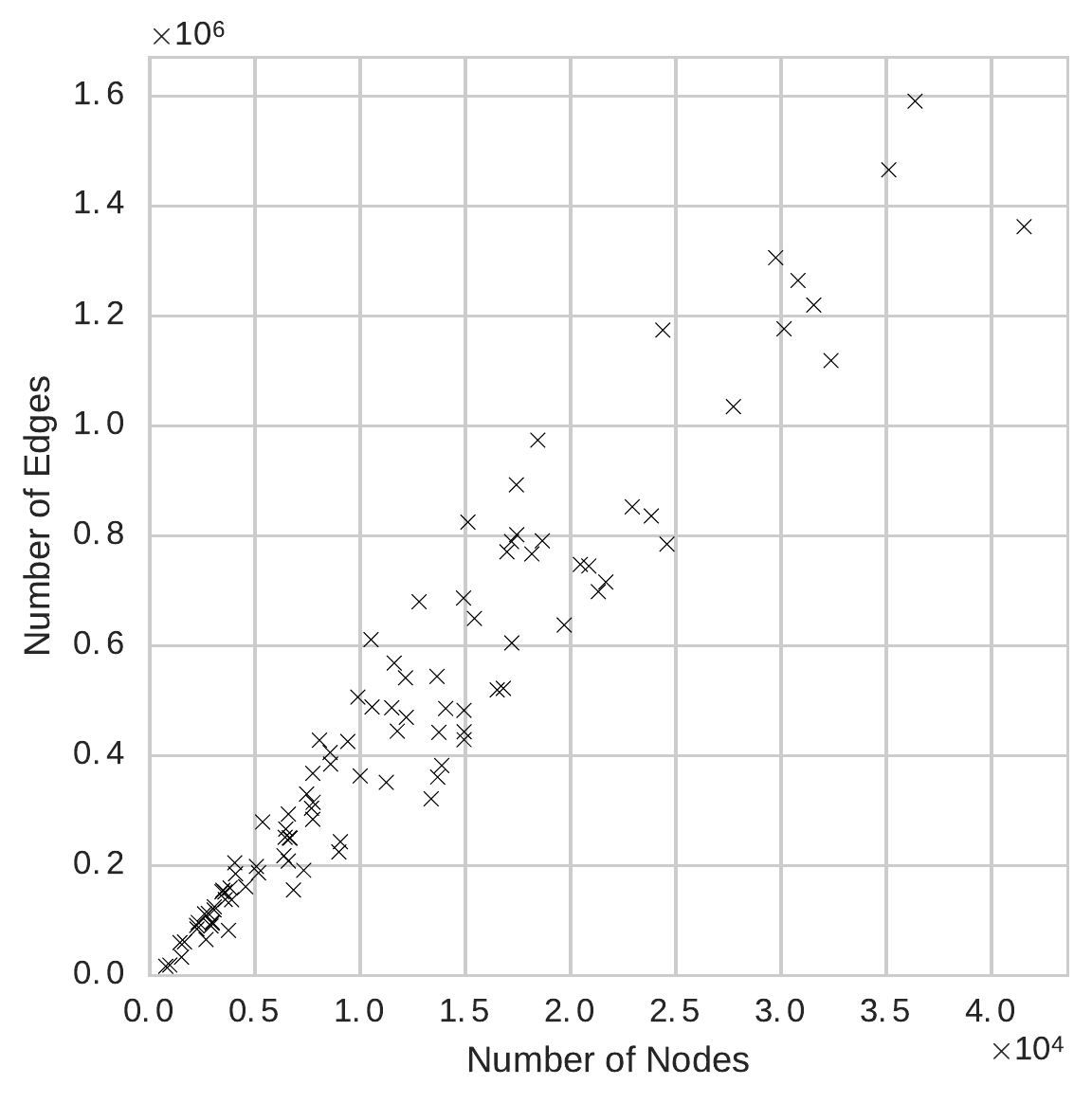}
\caption{Number of nodes and edges of the Facebook networks}
\label{fig:nodes_edges_fb}
\end{figure}

Our experiments have been performed on a multicore compute server with 4 physical Intel Core i7 cores at 3.4 GHz, 8 threads, and 32 GB of memory.
For this explorative study, we use a collection of 100 social networks representing early snapshots of Facebook, each of which is an online friendship network for a US university or college~\cite{traud2012social}, most members are students.
Sizes of the Facebook networks are between 10k and 1.6 million edges, the number of nodes and edges is shown in Fig.~\ref{fig:nodes_edges_fb}.
Unless otherwise noted, we aggregate experimental results over this set of networks.
The common origin and the high structural similarity among the networks allows us to get meaningful aggregated values.

For the experiments on the preservation of properties we also use the Twitter and Google+ networks \cite{leskovec2012circles} and the LiveJournal (com-lj) network from \cite{Yang:2012fk}.
All of them are friendship networks, the Twitter and Google+ networks consist of the combined ego networks of 973 and 132 users, respectively.
In Table~\ref{tab:network_stats} we provide the number of nodes and edges as well as diameter and clustering coefficient averaged over all Facebook networks and the individual values for the three other networks.
Furthermore we provide the number of edges divided by the number of nodes, which indicates how much redundancy there is in the network.
If this number was near 1 and the network connected, the network would be close to a tree in structure.
It is not realistic to expect that we can preserve the structure of the network if it is very sparse already.
The networks we selected have a varying degree of redundancy but all of them are dense enough such that if we remove 80\% of the edges more edges remain than we would need for a single tree.
The characteristics of the Facebook networks are relatively similar, we provide the individual values in the Appendix.

\begin{table}
\centering
\caption{Number of nodes $n$, the number of edges $m$, $m/n$, the diameter $D$ and the average local clustering coefficient $Cc$ of the used social networks (average values for the Facebook networks)}
\label{tab:network_stats}
\begin{tabular}{lrrrrr}
\toprule
{} &  $n$ &  $m$ & $m/n$ & $D$ & $Cc$ \\
Network   &           &           & &           &            \\
\midrule
Facebook &    12\,083 &    469\,845 &      38.4 &       7.8 &       0.25 \\
com-lj  &   3\,997\,962 &  34\,681\,189 &       8.7 &        21 &      0.35 \\
gplus   &    107\,614 &  12\,238\,285 &     113.7 &         6 &      0.52 \\
twitter &     81\,306 &   1\,342\,296 &      16.5 &         7 &      0.60 \\
\bottomrule
\end{tabular}
\end{table}

For the evaluation of the preservation of the community structure we also use networks with known ground truth communities.
For this purpose we use the synthetic LFR benchmark~\cite{PhysRevE.80.016118}.

It remains an open question to what extent results can be translated to other types of complex networks, since according to experience the performance of network analysis algorithms depends strongly on the network structure.

\subsection{Correlations between Edge Scores}
\label{sub:correlation}

\begin{figure*}[htb]
\begin{center}
\includegraphics[width=0.8\textwidth]{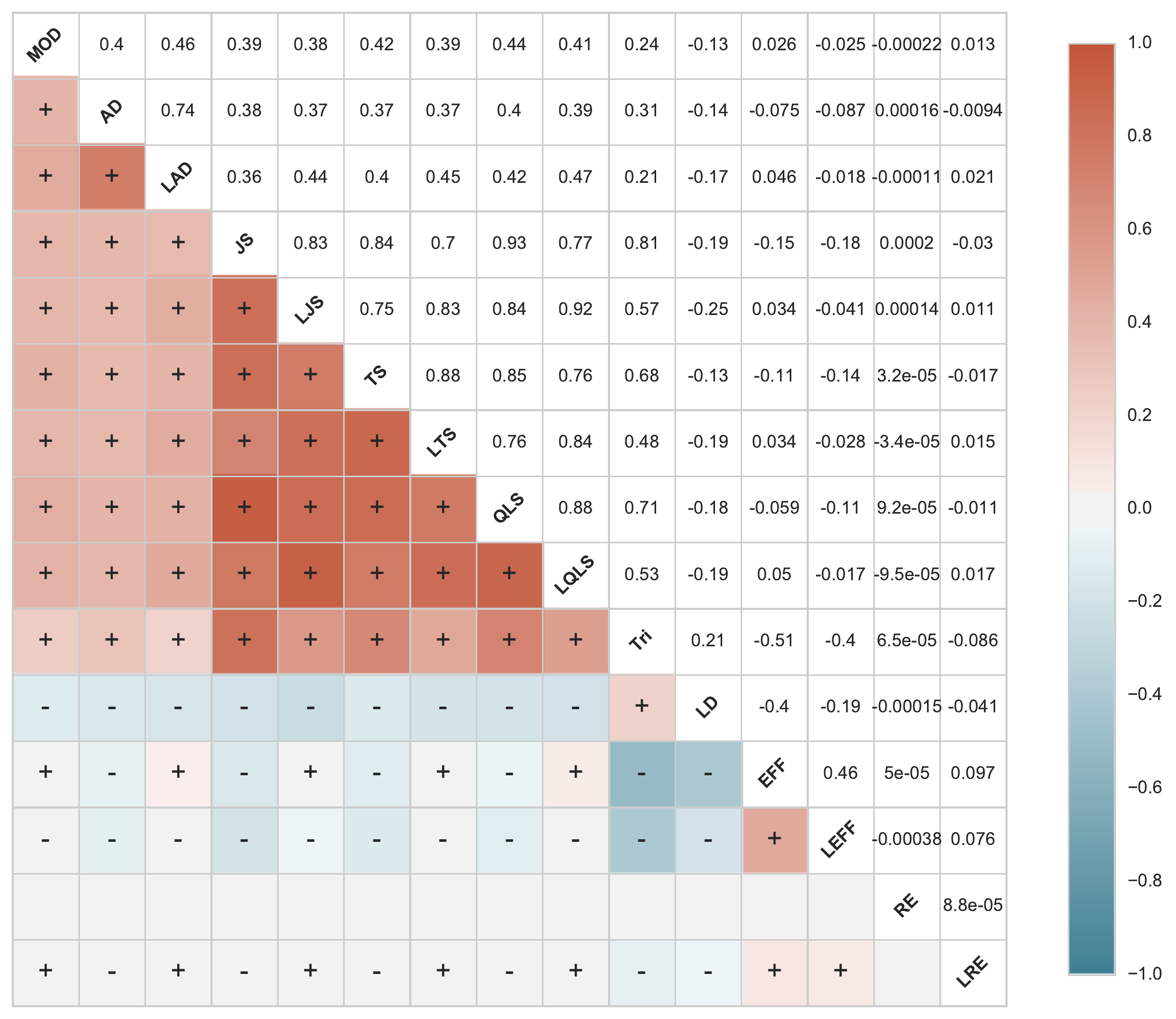}
\caption{Edge score correlations (Spearman's $\rho$, average over 100 Facebook networks)}
\label{fig:corrFb}
\end{center}
\end{figure*}

Among our sparsification methods, some are more similar to others in the sense that they tend to preserve similar edges. 
Such similarities can be clarified by studying correlations between edge scores.
We calculate edge score correlations for the set of 100 Facebook networks as follows:
For each single network, edge scores are calculated with the various scoring methods and Spearman's rank correlation coefficient is applied. 
The coefficient is then averaged over all networks and plotted in the correlation matrix (Fig.~\ref{fig:corrFb}).
There is one column for each method, and the column \textsf{Mod} represents edge scores that are 1 for intra-community edges and 0 for inter-community edges after running a modularity-maximizing Louvain community detection algorithm.
Positive correlations with these scores indicate that the respective rating method assigns high scores to edges within modularity-based communities. 
The column \textsf{Tri} simply represents the number of triangles an edge is part of.
As some of the methods are normalizations of this score, this shows how similar the ranking still is to the original score.

Interpretation of the results is challenging: 
The correlations we observe reflect intrinsic, mathematical similarities of the rating algorithms on the one hand, but on the other hand they are also caused by the structure of this specific set of social networks (e.g., it may be a characteristic of a given network that edges leading to high-degree nodes are also embedded in many triangles).
Nonetheless, we note the following observations:
There are several groups of methods.
Simmelian Backbones, Jaccard Similarity and Triangles are highly positively correlated which is not unexpected as they are all based on triangles or quadrangles and are intended to preserve dense subgraphs.
Algebraic distance is still positively correlated with these methods but not as strongly even though they are also intended to prefer dense subgraphs.
An explanation for this weaker correlation is that while both prefer dense regions, the order of the individual edges is different.
All of the previously mentioned methods are also correlated with the Modularity value, algebraic distance with local filtering has the highest score among all of these methods.
Our experiments on the preservation of community structure (Sec. \ref{sec:quantifying}) confirm this relationship.
The correlation of the Modularity value and these methods are similar to the correlation between algebraic distance and the rest of the methods which shows again that the lower correlation values are probably due to different orderings of the individual edges.

Our new method Local Degree is slightly negatively correlated with all these methods but still positively correlated with the Triangles.
It is also slightly negatively correlated with the Modularity value, this is due to the method's preference of inter-cluster edges which is also confirmed by our experiments below.
The newly introduced Edge Forest Fire is also negatively correlated with Local Degree and even more negatively with Triangles.
This strong negative correlation between Edge Forest Fire and triangle count can be explained by the fact that the Edge Forest Fire can never ``burn'' a triangle, as nodes cannot be visited twice.
Random edge filtering is not correlated at all, which is definitely expected.

It is interesting to see that each method is also relatively strongly correlated with its local variant, apart from random edge filtering (we use different random values as basis of the local filtering process).
Even the Edge Forest Fire method, which should also be relatively random, has a positive correlation with its local variant. This shows that it prefers a certain kind of edge and that this preference is kept when applying the local filtering.

Among the variants of Simmelian Backbones and Jaccard Similarity also the local variants are more correlated to other local variants than to other non-local variants and also not as strongly correlated to triangles.
This shows that the local filtering indeed adds another level of normalization.
Also Jaccard Similarity seems to be more correlated to Quadrilateral Simmelian Backbones than to the variant based on triangles even though Jaccard Similarity is based on triangles itself.
This is also interesting to see, as Quadrilateral Simmelian Backbones are computationally more expensive than the Jaccard Similarity.

\subsection{Similarity in Network Properties}
\label{sec:quantifying}

Quantifying the similarity between a network and its sparsified version is an intricate problem.
Ideally, a similarity measure should meet the following requirements:
\begin{enumerate}
\item \emph{Ignoring trivial differences}: Consider, for example, the degree distribution: One cannot expect the distribution to remain identical after edges get removed during sparsification.
It is clear, however, that the general shape of the distribution should remain ``similar'' and that high-degree nodes should remain high-degree nodes in order to consider the degrees as preserved.
\item \emph{Intuitive and Normalized}: Similarity values from a closed domain like $[0,1]$ allow for aggregation and comparability.
A similarity value of $1$ indicates that the property under consideration is fully preserved, whereas a value of $0$ indicates that similarity is entirely lost.
In some cases we also used relative changes in the interval $[-1,1]$ where $0$ means unchanged as they provide a more detailed view at the changes.
\item  \emph{Revealing Method Behavior}: A good similarity measure will clearly expose different behavior between sparsification methods.
\item  \emph{Efficiently computable}.
\end{enumerate}

Following these requirements, we select measures that quantify relative changes for global properties like diameter, size of the largest connected component and quality of a community structure.
Node degree, betweenness and PageRank can be treated as node centrality indices which represent a ranking of nodes by structural importance.
Since absolute values of the centrality scores are less interesting than the resulting rank order,
we compare the rankings before and after sparsification using Spearman's $\rho$ rank correlation coefficient.
(This focus on rank order is also the reason why we did not adopt the Kolmogorov-Smirnov statistic used in~\cite{Leskovec2006}, which compares distributions of absolute values.)
Even though the local clustering coefficient can be interpreted as a centrality score as well, the comparison of ranks does not seem meaningful in this case due to the fact that it is a local score.
Instead, we analyse the deviation of the average local clustering coefficient from the original value.

In the following plots, the measures are shown on the y-axis for a given ratio of kept edges  ($m' / m$) on the x-axis (e.g., a ratio of 0.2 means that 20\% of edges are still present).
For each value there are two rows of plots.
The first contains averages over the 100 Facebook networks with error bars that indicate the standard deviation.
The second row contains the values at 20\%, 50\% and 80\% remaining edges of the three other networks.
In each row, we show two plots: the left plot with the non-local methods and the right plot with the methods that use local filtering.

\paragraph*{Connected Components. \,}

\begin{figure}[htbp]
\begin{center}

\includegraphics[width=0.5\columnwidth]{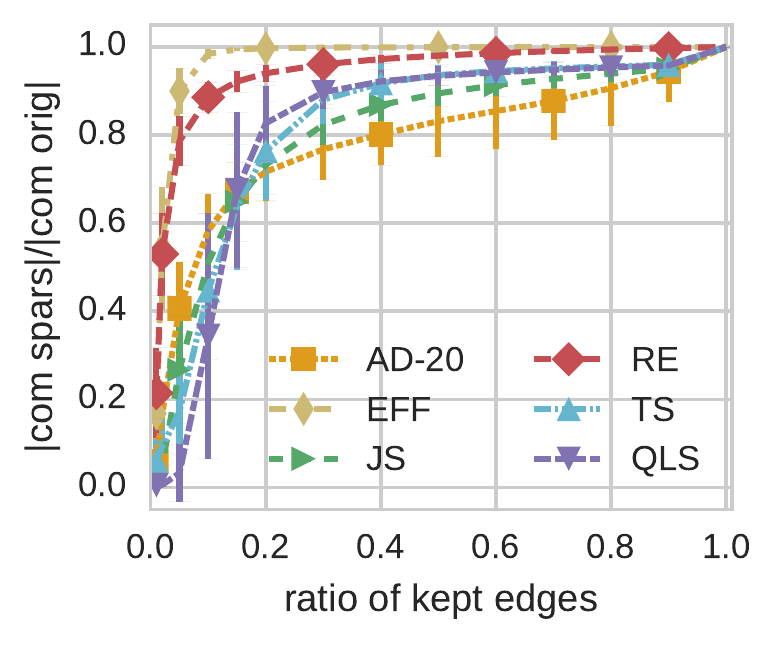}%
\includegraphics[width=0.5\columnwidth]{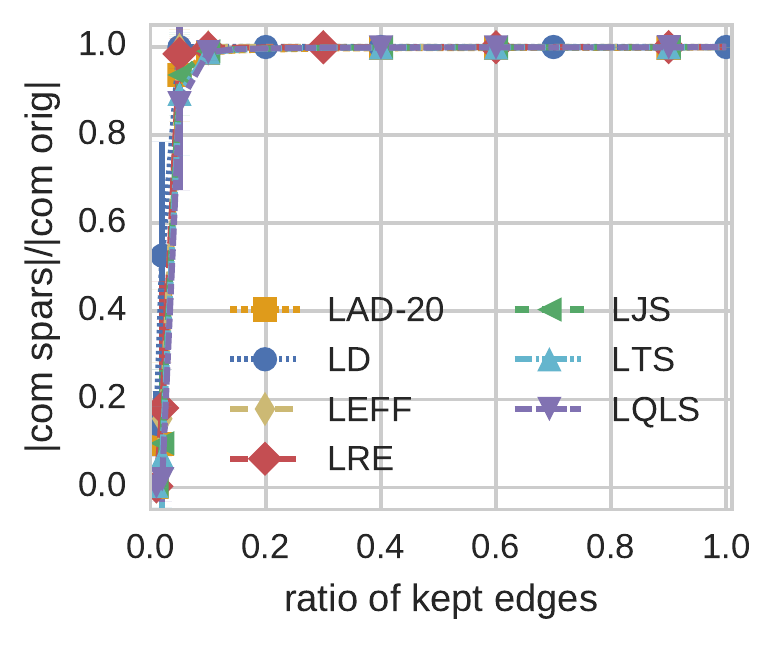}

\includegraphics[width=0.5\columnwidth]{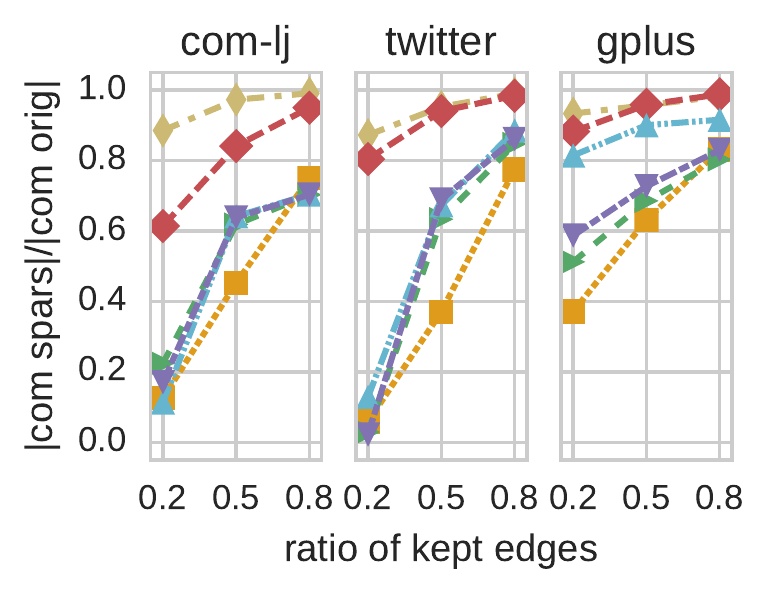}%
\includegraphics[width=0.5\columnwidth]{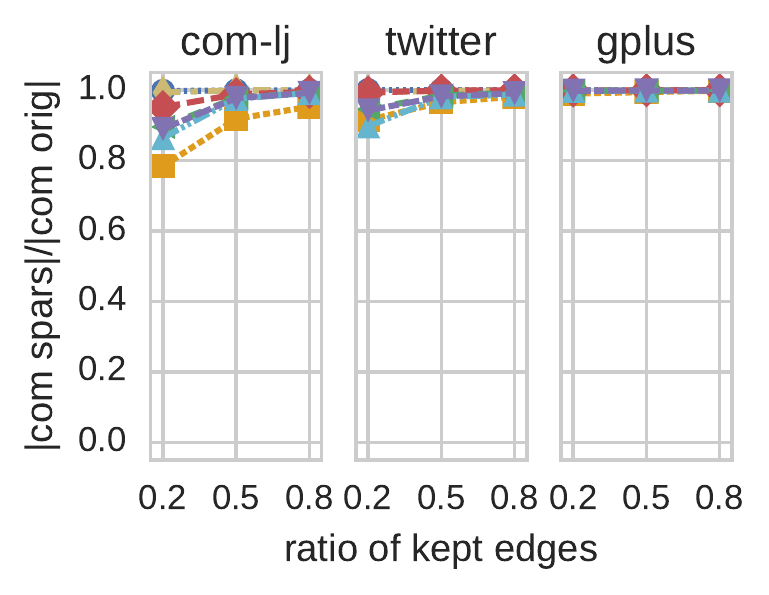}
\caption{The size of the largest component in the sparsified network divided by the size of the largest component in the original network.}
\label{fig:plt_wcc_factor}

\end{center}
\end{figure}

As all of our networks have, like most real-world complex networks, a giant component that comprises most nodes we track its change by dividing the size of the largest component in the sparsified network by the size of the largest component in the original network.
As shown in Fig.~\ref{fig:plt_wcc_factor}, out of the non-local methods Edge Forest Fire best preserves the connected component.
Random edge deletion leads to a slow decrease in the size of the largest component while Simmelian Backbones, Jaccard Similarity and algebraic distance lead to a separation very quickly.
Below 20\% of retained edges, the size of the largest component on the Facebook networks drops very quickly, here the networks seem to be decomposed into multiple smaller parts.
On the other networks, this drop occurs at different ratios of kept edges which reflects their different densities and probably also their different structures.
Local Filtering is able to maintain the connectivity.
On the Facebook networks, all methods keep the largest component almost fully connected up to 20\% of retained edges, only below that small differences are visible.
The results on the LiveJournal, Twitter and Google+ networks show that -- as expected -- with increasing density it is easier to preserve the connectivity of the network.
Our Local Degree method best preserves the connected components of all networks, closely followed by the local variant of random edge deletion and Edge Forest Fire.

\paragraph*{Diameter. \,}

In order to observe how the network diameter changes through sparsification, we plot the quotient of the original network diameter and the resulting diameter, which yields legible results since in practice the diameter is mostly increased during sparsification.
We compute the exact diameters using a variation of the ExactSumSweep algorithm \cite{Borassi2015}.

\begin{figure}[htb]

\begin{center}

\subfloat[Original network diameter divided by network diameter ]{%
\begin{minipage}{\columnwidth}%
\includegraphics[width=0.5\columnwidth]{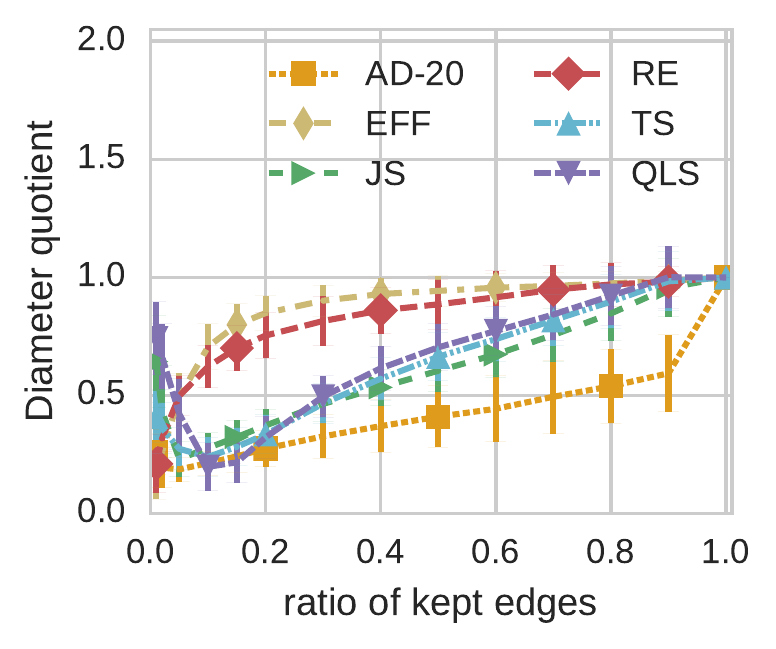}%
\includegraphics[width=0.5\columnwidth]{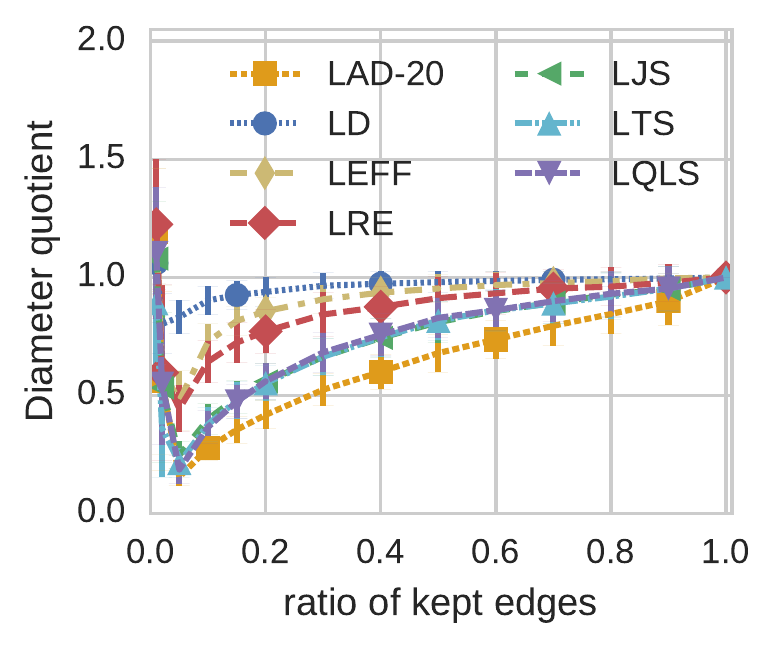}
\includegraphics[width=0.5\columnwidth]{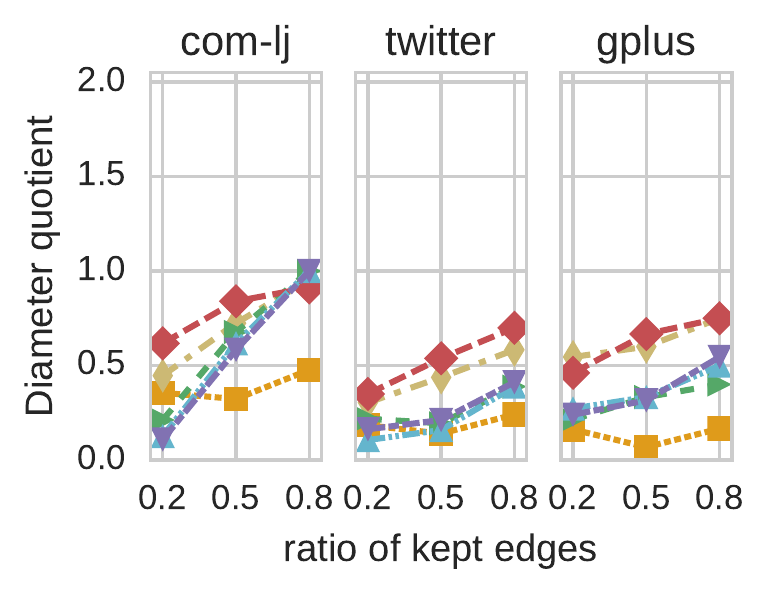}%
\includegraphics[width=0.5\columnwidth]{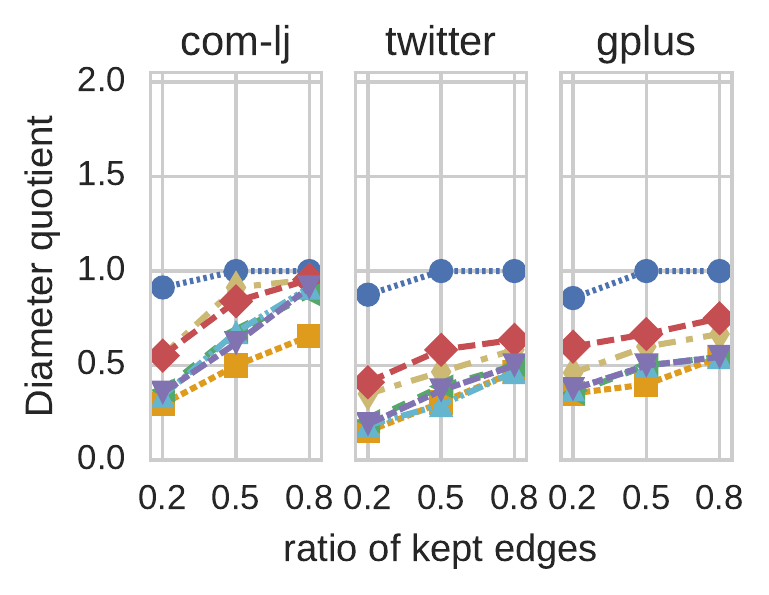}%
\end{minipage}%
\label{fig:plt_diameter}%
}

\subfloat[Deviation from original clustering coefficient]{%
\begin{minipage}{\columnwidth}%
\includegraphics[width=0.5\columnwidth]{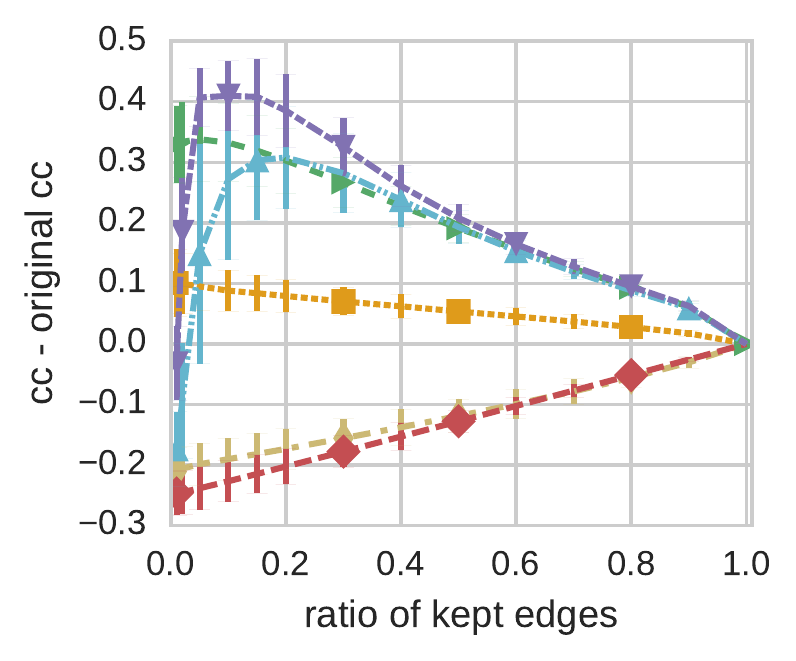}%
\includegraphics[width=0.5\columnwidth]{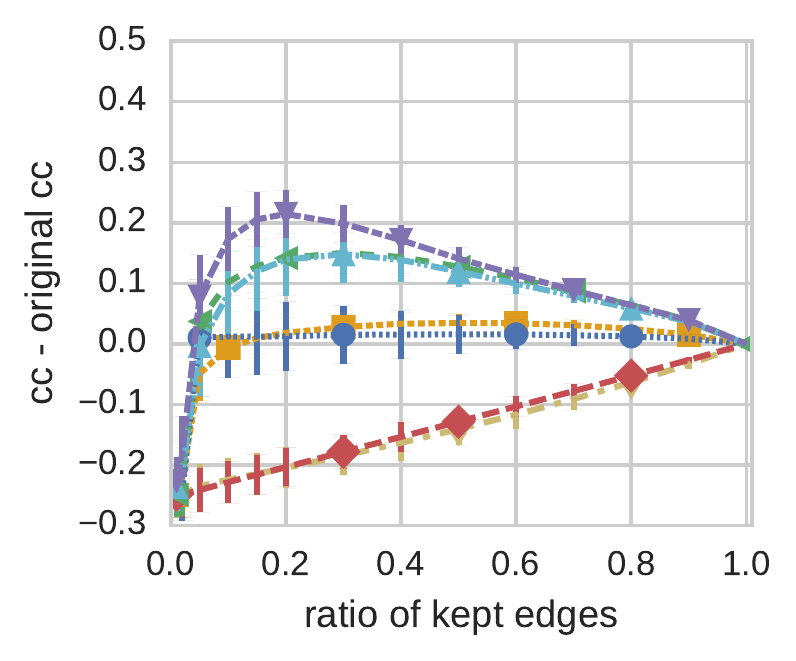}
\includegraphics[width=0.5\columnwidth]{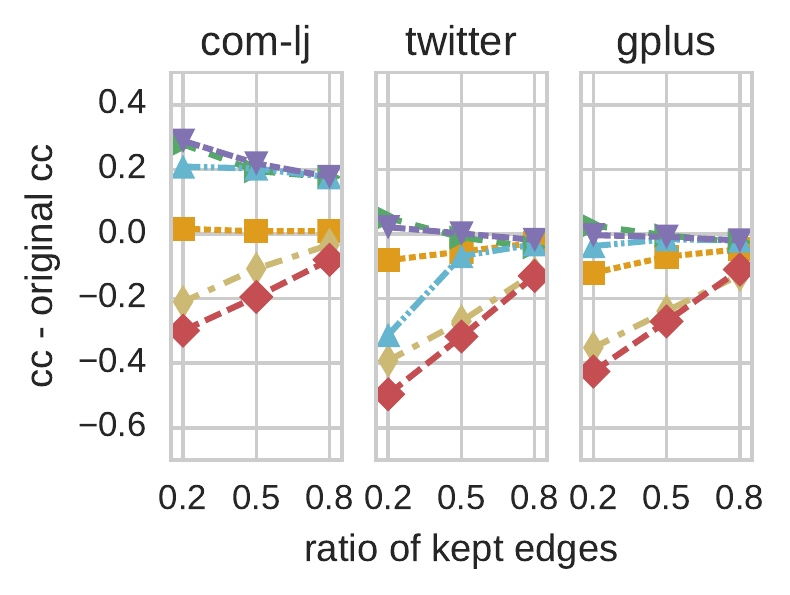}%
\includegraphics[width=0.5\columnwidth]{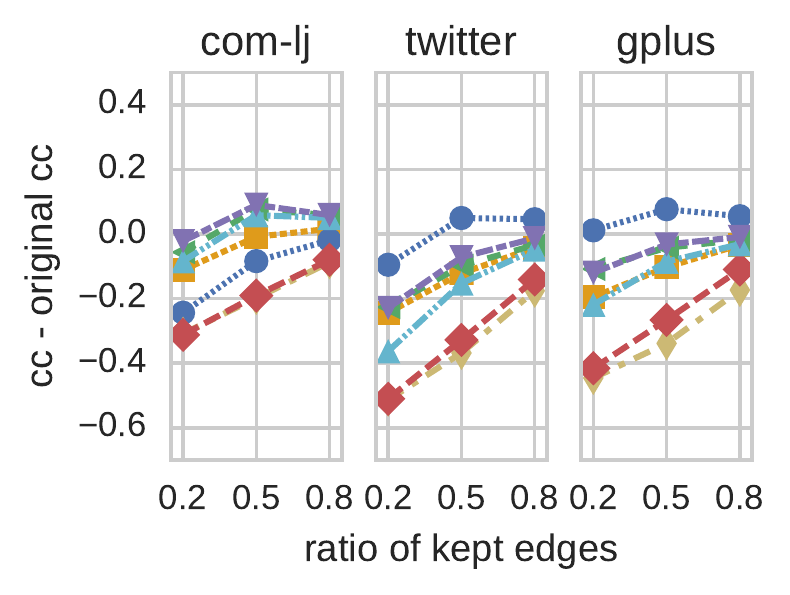}%
\end{minipage}%
\label{fig:plt_ccAvgLocal}%
}

\caption{Preservation of global network properties}

\end{center}
\end{figure}

We motivate the Local Degree method with the idea that shortest paths commonly run through hub nodes in social networks.
Therefore, preserving edges leading to high-degree nodes should preserve the small diameter.
This is confirmed by our experiments (Fig.~\ref{fig:plt_diameter}).
In contrast, methods that prefer edges within dense regions clearly do not preserve the diameter.
With Simmelian Backbones the diameter drops when only few edges are left; this can be explained by the fact that Simmelian Backbones do not maintain the connectivity and that at the end the graph is decomposed into multiple connected components which have a smaller diameter.
Algebraic distance is even more extreme in this aspect.
Local filtering leads to a slightly better preservation of the diameter when applied to the other methods but algebraic distance remains the worst method in this regard.
Note that the LiveJournal network has a higher diameter than the other networks (see Table~\ref{tab:network_stats}); this might explain why the diameter is better preserved there.

\paragraph*{Clustering Coefficient. \,}

Fig.~\ref{fig:plt_ccAvgLocal} shows the deviation of the average local clustering from the value of the original network.
Both for local and non-local methods we observe three classes of methods on the Facebook networks: methods that clearly decrease the clustering coefficient, methods that preserve the clustering coefficient and methods that increase it.

For both Random Edge and Edge Forest Fire, which are based on randomness, the clustering coefficient drops almost linearly with decreasing sparsification ratio.
This can also be observed on the other three networks.
The additional local filtering step does not significantly change this.

Simmelian Backbones and Jaccard Similarity keep mostly edges within dense regions, which results in increasing clustering coefficients on all networks.
Triadic Simmelian Backbones show the weakest increase, on the Twitter network even a decrease of the clustering coefficients.
Note that with 0.52 and 0.6 the clustering coefficients are already relatively high on the Google+ and Twitter networks, therefore the very small increase is not surprising.
Local filtering slightly weakens this effect on the Facebook networks, on the other networks it is even reversed.
Given the high clustering coefficients in the original networks, this is not very surprising as we would need to retain very dense areas while local filtering leads to a more balanced distribution of the edges.

From the previous results especially concerning the connected components one would expect that algebraic distance also increases the clustering coefficients.
Interestingly though, filtering using algebraic distance leads to a slight increase of the clustering coefficient on the Facebook networks, constant clustering coefficients on the LiveJournal network and even slightly decreasing clustering coefficients on the Twitter and Google+ networks.
With the additional local filtering step algebraic distance almost preserves the clustering coefficients on the Facebook networks while on the other networks it is slightly decreased.
Algebraic distance leads to random noise on the individual edge weights, therefore they probably lead to a more random selection of edges that also destroys more triangles than the selection of Simmelian Backbones and Jaccard Similarity.
Our Local Degree method best preserves the clustering coefficient on the Facebook networks, though with some differences between the various networks in the dataset (note the error bars).
On the LiveJournal network it leads to a decrease of the clustering coefficient while on the Twitter and Google+ networks it leads to a slight increase of the clustering coefficient.
This is probably due to the special structure of ego networks.

Our experiments therefore do not reveal a general best method for preserving clustering coefficients.
If high clustering coefficients shall be created or preserved, the Jaccard Similarity and Quadrilateral Simmelian Backbones seem to be a good choice.
Algebraic distance is good at preserving clustering coefficients with slight deviations but our Local Degree method also works well on the considered social networks.

\paragraph*{Node Centrality Measures. \,}

The exact calculation of betweenness centrality is in practice too expensive for the whole set of networks and sparsification methods we consider.
Therefore we use the approximation algorithm \cite{geisberger2008better} with at least 16 samples, for smaller networks also with up to 512 samples.
For the calculation of the PageRank centrality we use a damping factor of 0.85 and an error tolerance of $10^{-9}$.

\begin{figure}[tbp]

\subfloat[Preservation of node degree ]{%
\begin{minipage}{\columnwidth}%
\includegraphics[width=0.5\columnwidth]{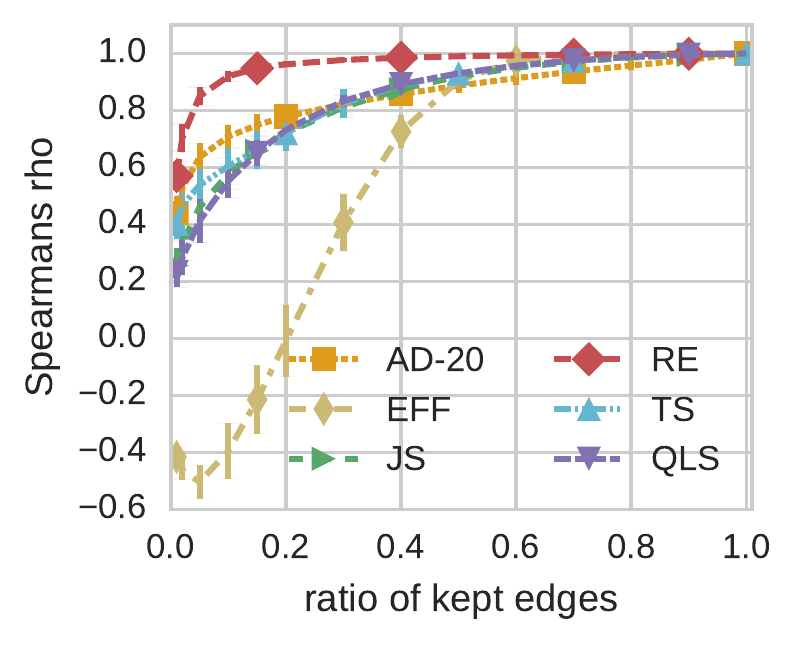}%
\includegraphics[width=0.5\columnwidth]{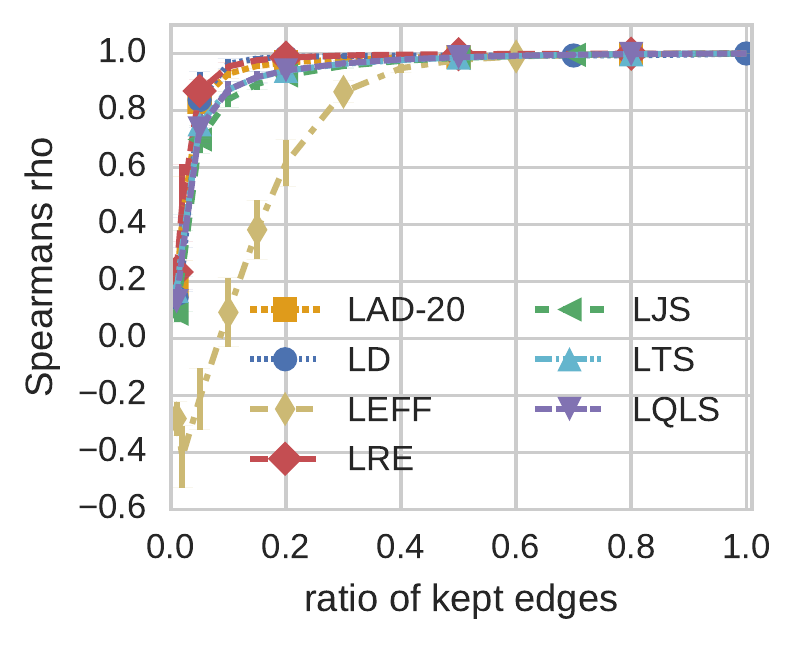}
\includegraphics[width=0.5\columnwidth]{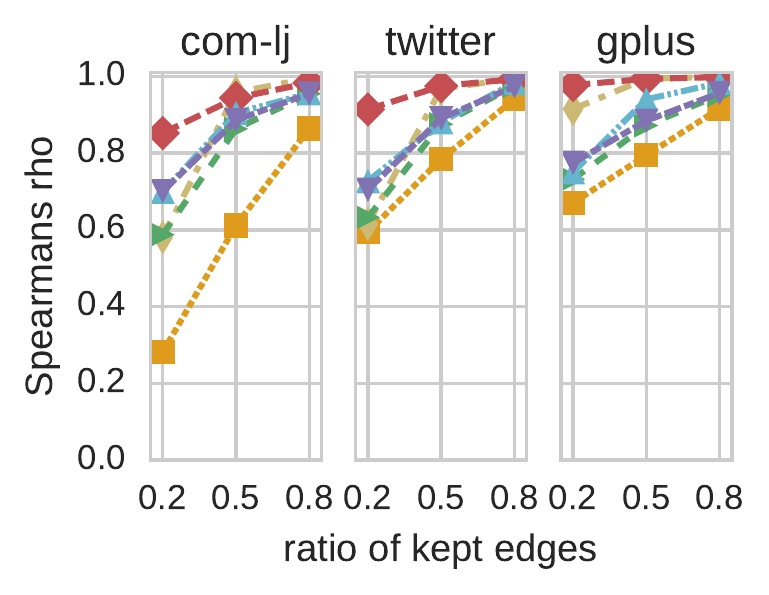}%
\includegraphics[width=0.5\columnwidth]{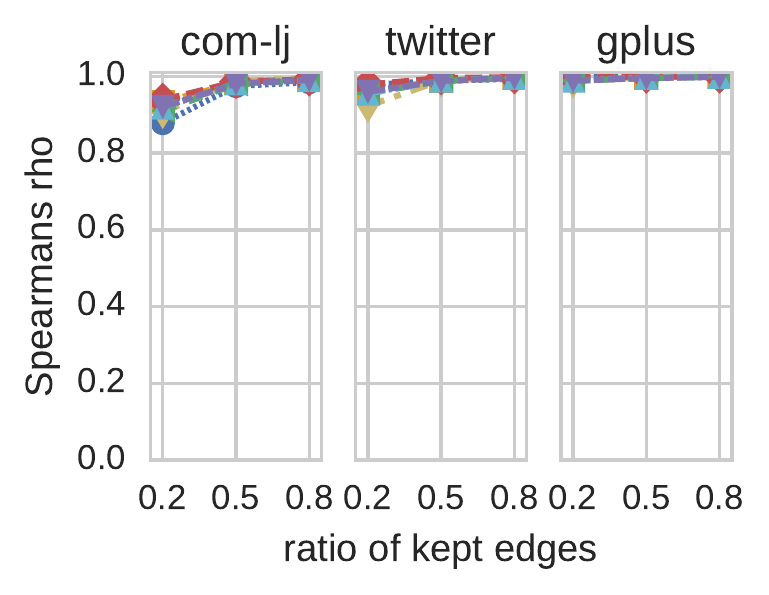}%
\end{minipage}%
\label{fig:plt_dd_spearman_rho}%
}

\subfloat[Preservation of betweenness centrality ]{%
\begin{minipage}{\columnwidth}%
\includegraphics[width=0.5\columnwidth]{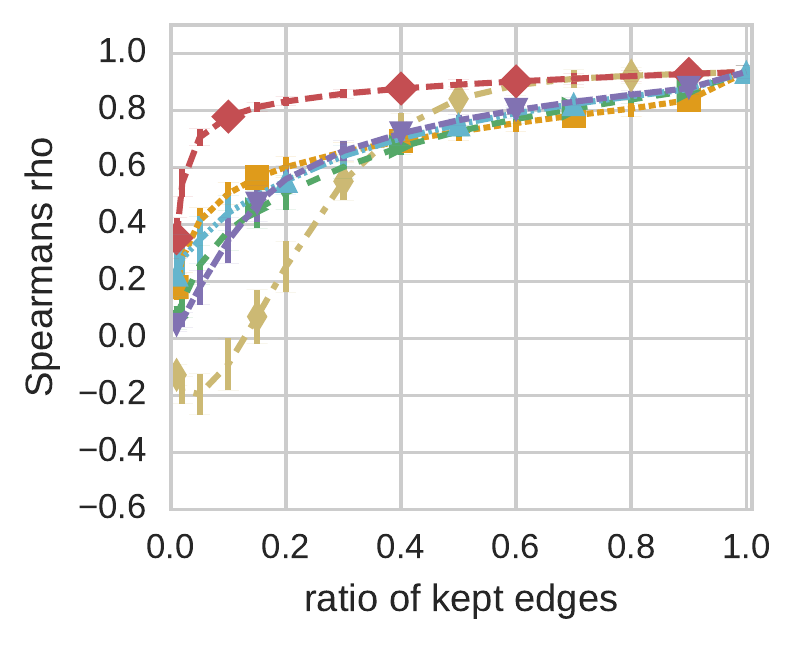}%
\includegraphics[width=0.5\columnwidth]{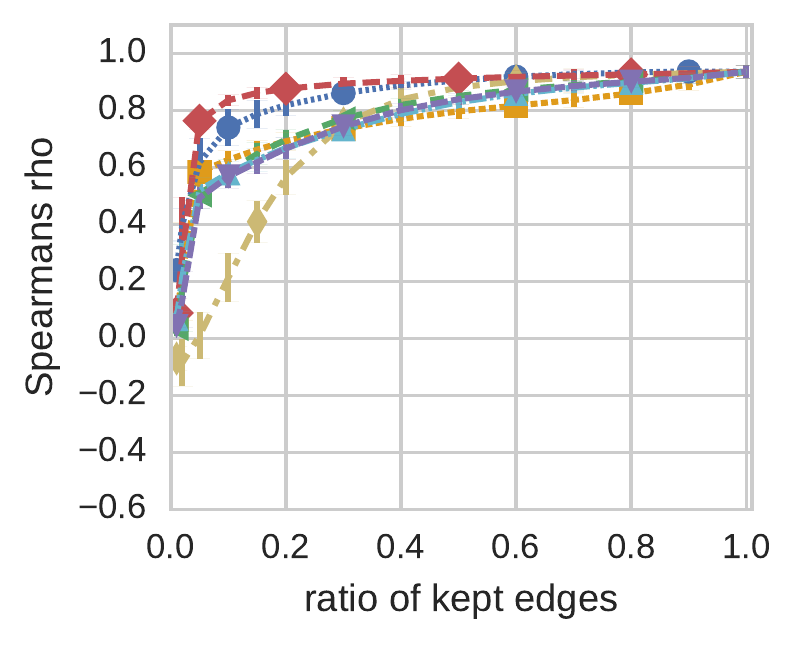}
\includegraphics[width=0.5\columnwidth]{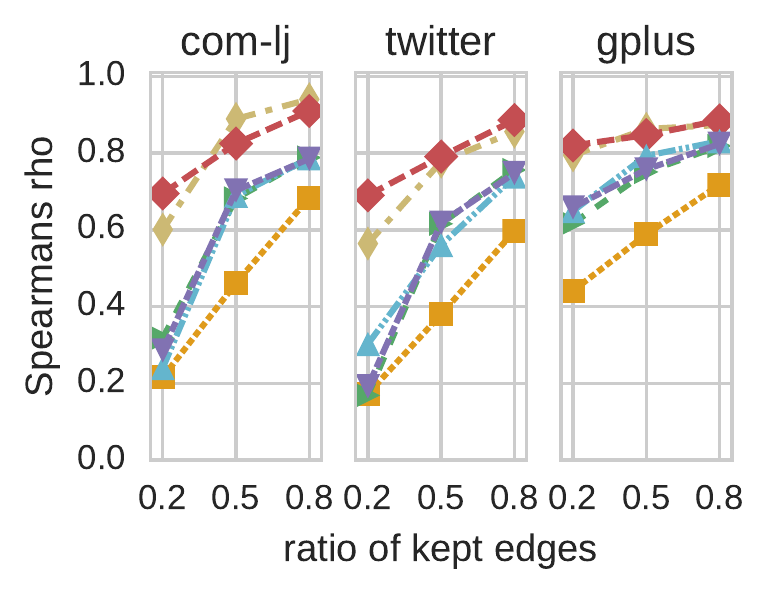}%
\includegraphics[width=0.5\columnwidth]{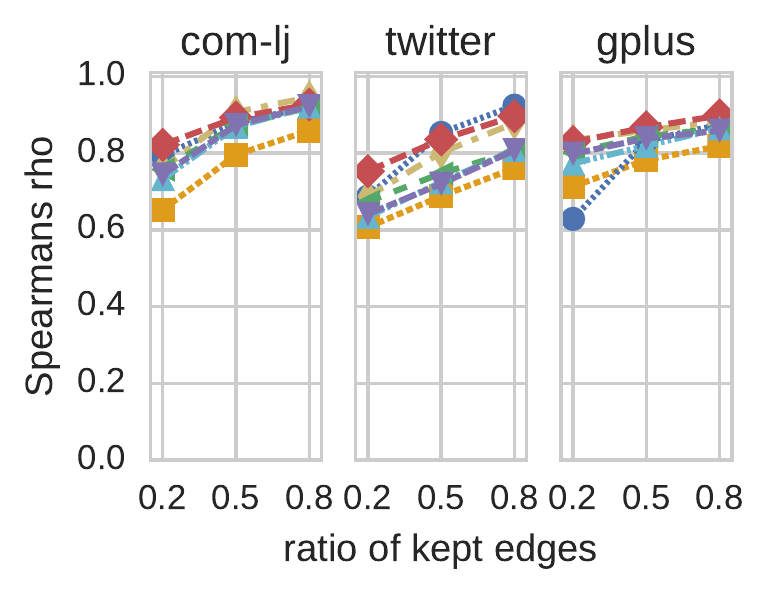}%
\end{minipage}%
\label{fig:plt_betweenness_spearman_rho}%
}

\subfloat[Preservation of PageRank centrality ]{%
\begin{minipage}{\columnwidth}%
\includegraphics[width=0.5\columnwidth]{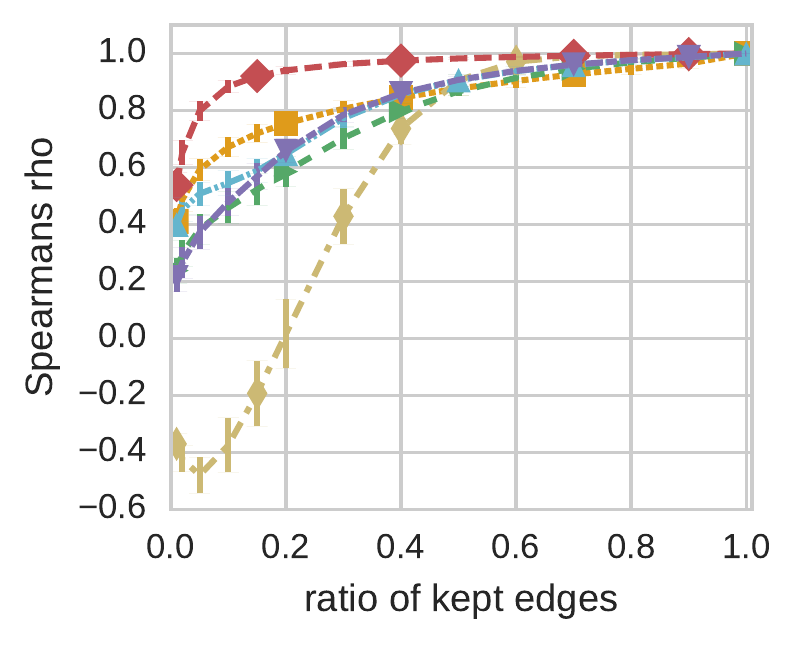}%
\includegraphics[width=0.5\columnwidth]{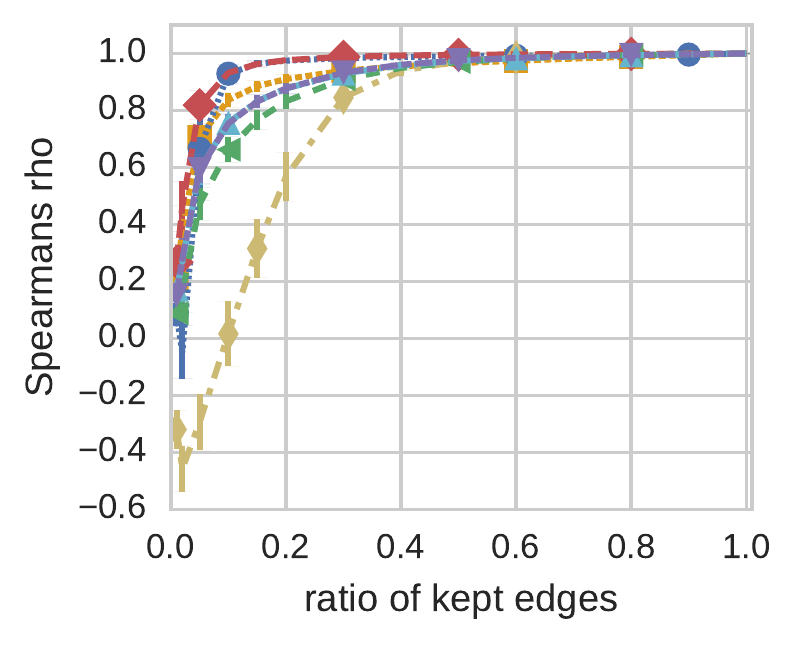}
\includegraphics[width=0.5\columnwidth]{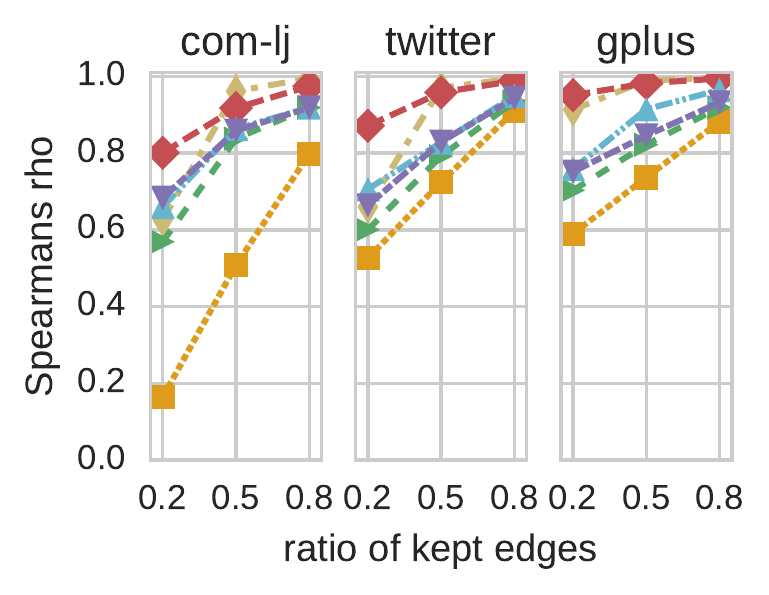}%
\includegraphics[width=0.5\columnwidth]{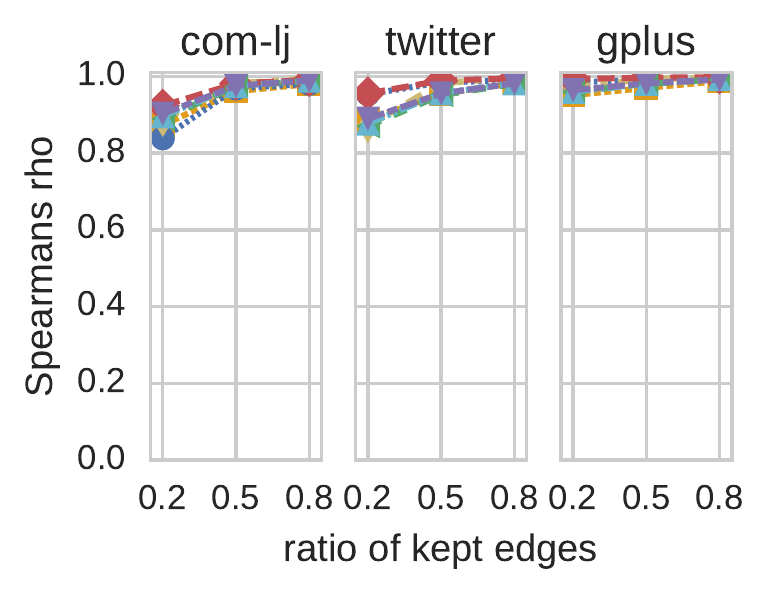}
\end{minipage}%
\label{fig:plt_pagerank_spearman_rho}%
}

\caption{Preservation of the ranking of node centrality measures (Spearman's $\rho$ rank correlation coefficient)}
\label{fig:centralities}
\end{figure}

The similarity of curves in Fig.~\ref{fig:centralities} catches the eye immediately:
For these node centrality measures, the sparsification methods behave in a very similar way. 
This similarity could be explained by strong correlations between node degree, PageRank and betweenness, which have been observed before (e.g.~\cite{fortunato2008approximating}).
Note that betweenness centrality is also not exactly preserved on the original network; this is due to the approximation, which adds additional noise.

Random edge deletion and Local Degree perform best on most networks.
In accordance with our intuition that edges leading to high-degree neighbors are important and should be preserved, our experiments show that the Local Degree method preserves all three considered node centralities. Nevertheless, random edge filtering with the additional local filtering step outperforms it concerning the preservation of Betweenness Centrality.
The differences are small though and similar to those that are due to the approximation error so they might actually be caused by the approximation method for Betweenness centrality that behaves differently depending on the structure of the network.
On the Facebook networks, Edge Forest Fire fails early while on the other networks it is among the best methods.
As the expected number of randomly selected incident edges via the ``burning process'' of Edge Forest Fire is relatively low even for high-degree nodes, it fails at preserving node degrees.
Nevertheless in the non-Facebook networks it seems to preserve enough important connections in order to preserve PageRank and betweenness centralities relatively well.
Methods that are focused on keeping edges within dense regions are not as good at preserving these centralities.
Adding the additional local filtering step again leads to a better preservation of the properties but does not change the general picture.

\paragraph*{Community Structure. \,}

In order to understand how the community structure of the networks is maintained, we consider for each network a fixed community structure that has been found by the Louvain method on the original network.
We report some properties of this community structure for each level of sparsification.
There are many ways to characterize a community structure.
We pick two properties of communities that we consider to be crucial.
Communities are commonly described to be internally dense and externally sparse subgraphs.
A natural measure is thus \emph{conductance} which compares the size of the cut of a community to the volume of the community, i.e. the sum of all degrees (or the volume of the rest of the network if it should be larger).
Low conductance values indicate clearly separable communities.
We consider the average conductance of all communities.
Furthermore we expect that communities are connected.
In order to measure this, we introduce the fraction of the nodes in a community that does not belong to the largest connected component of the community as partition fragmentation.
We report the average fragmentation of all communities.

\begin{figure}[tbp]
\begin{center}

\subfloat[Relative conductance change of a fixed partition]{%
\begin{minipage}{\columnwidth}%
\includegraphics[width=0.5\columnwidth]{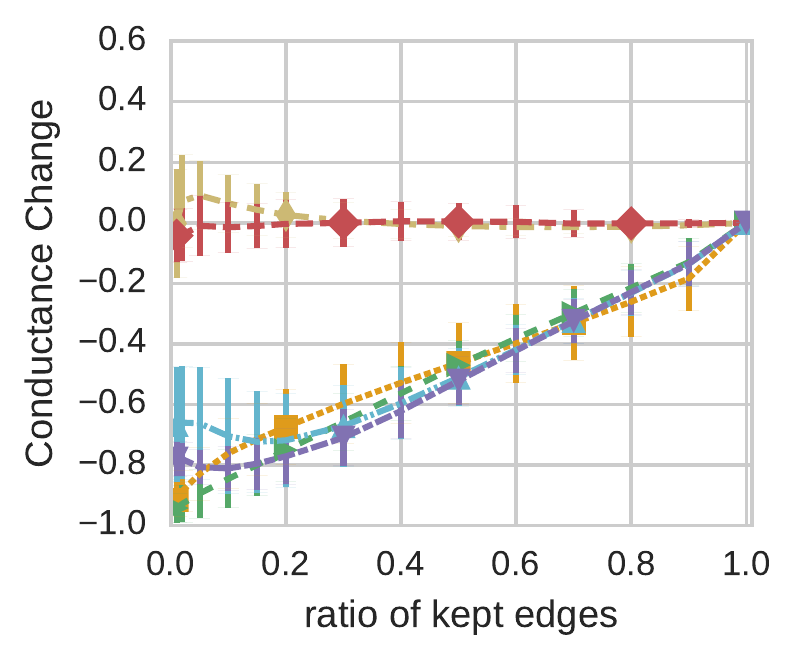}%
\includegraphics[width=0.5\columnwidth]{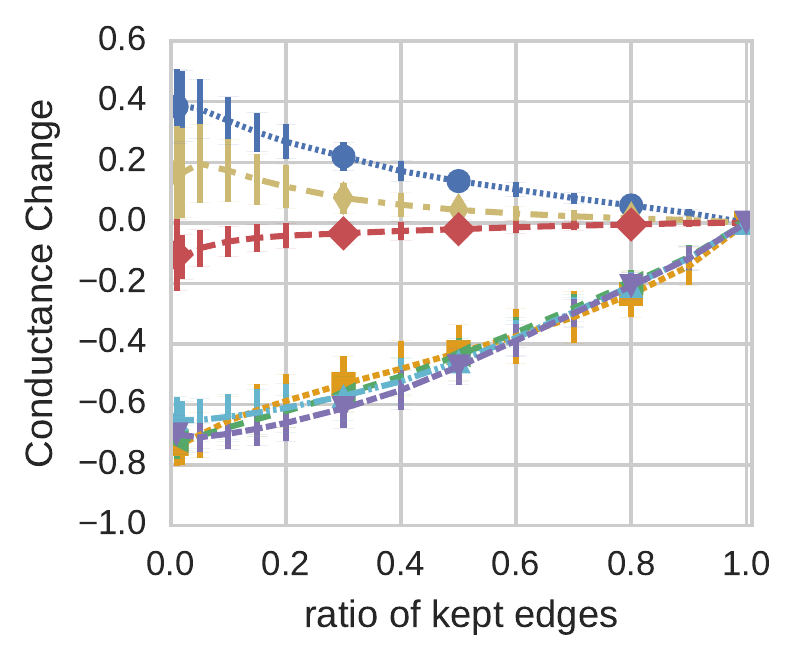}
\includegraphics[width=0.5\columnwidth]{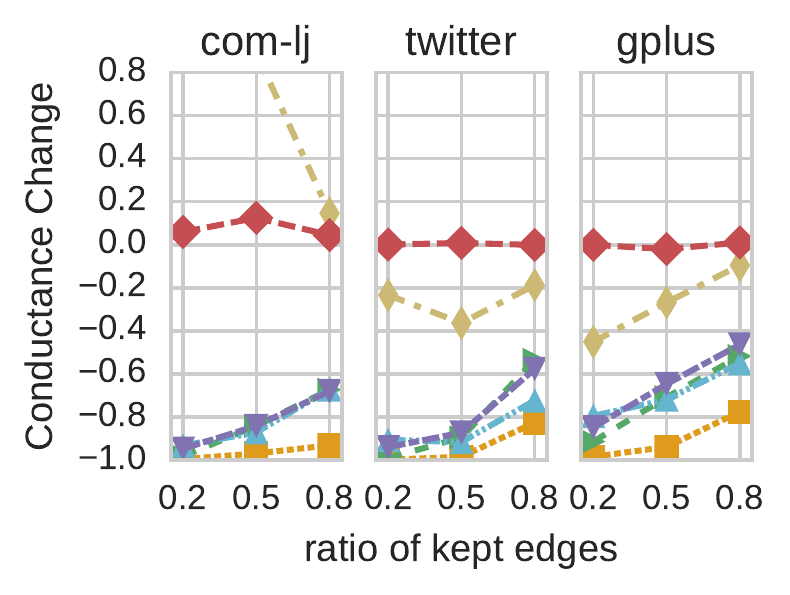}%
\includegraphics[width=0.5\columnwidth]{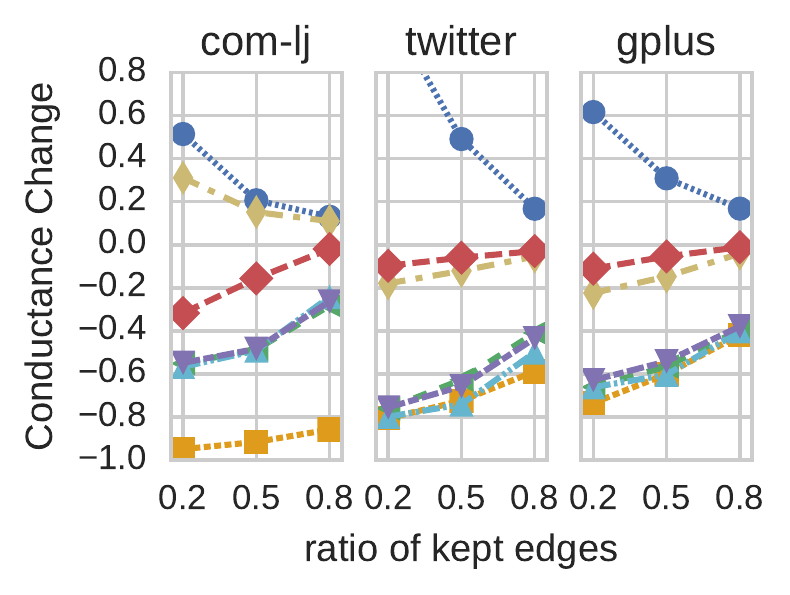}%
\end{minipage}%
\label{fig:plt_aixc_difference_percentage}%
}

\subfloat[Average partition fragmentation]{%
\begin{minipage}{\columnwidth}%
\includegraphics[width=0.5\columnwidth]{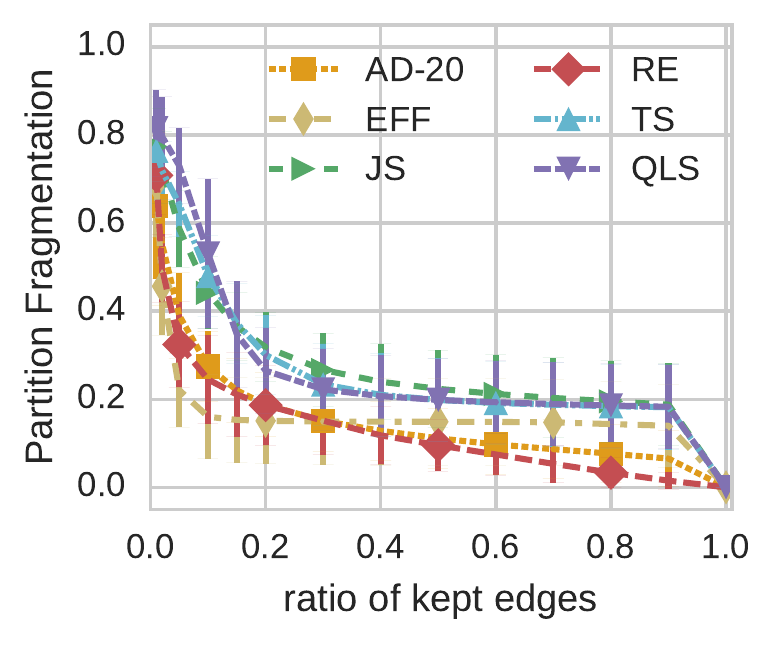}%
\includegraphics[width=0.5\columnwidth]{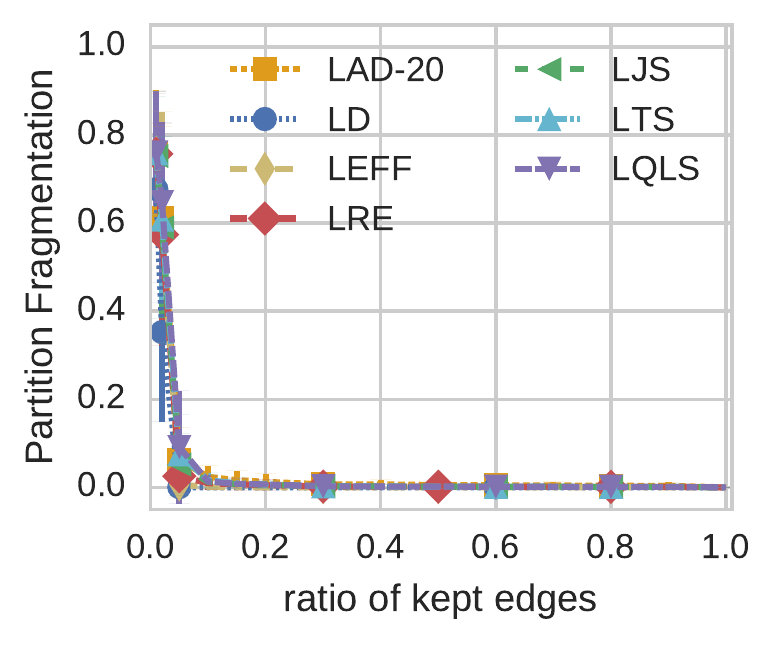}
\includegraphics[width=0.5\columnwidth]{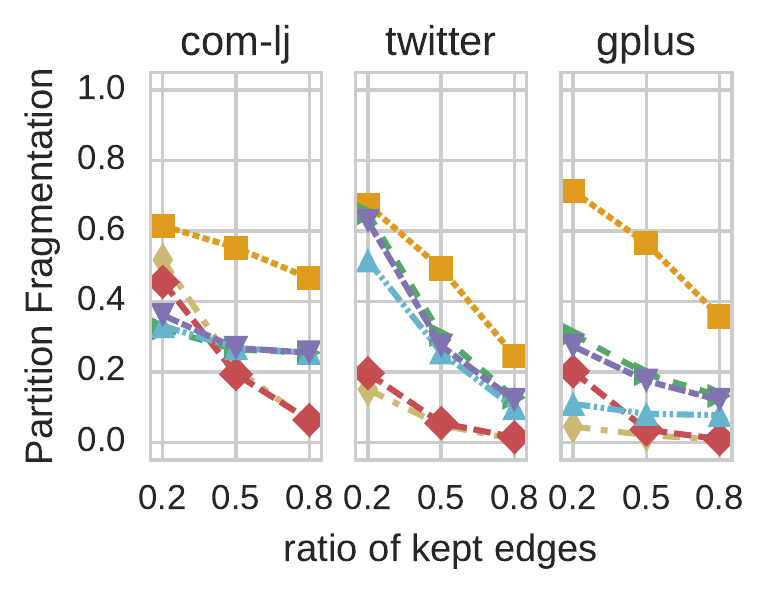}%
\includegraphics[width=0.5\columnwidth]{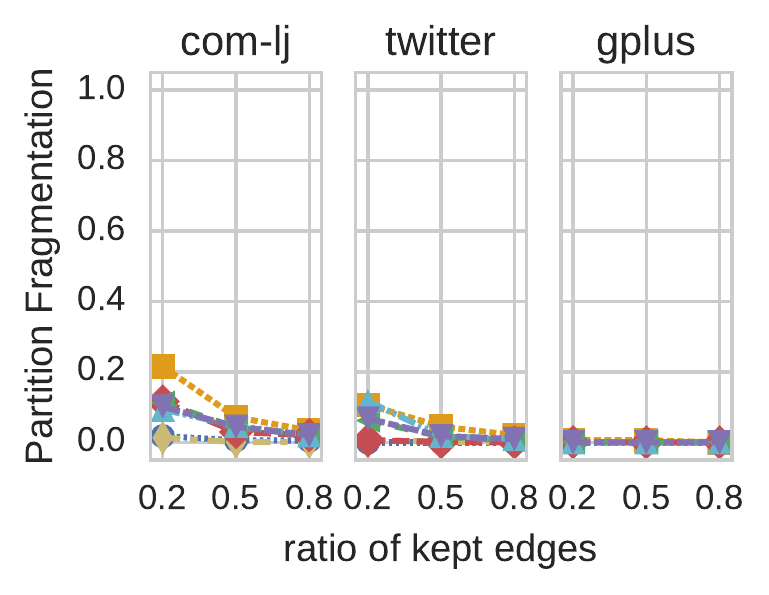}%
\end{minipage}%
\label{fig:plt_average_fragmentation}%
}

\subfloat[Adjusted rand measure between partition into communities on the original network and on the sparsified network]{%
\begin{minipage}{\columnwidth}%
\includegraphics[width=0.5\columnwidth]{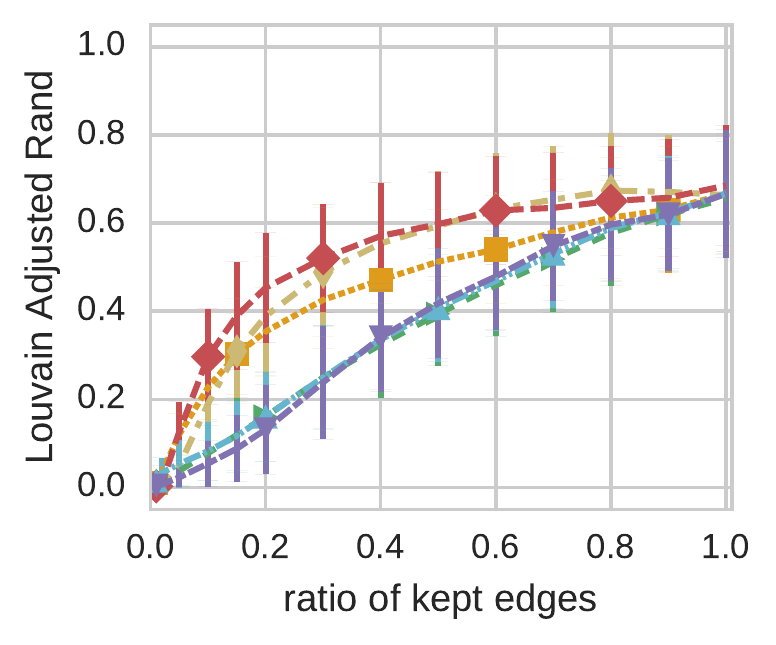}%
\includegraphics[width=0.5\columnwidth]{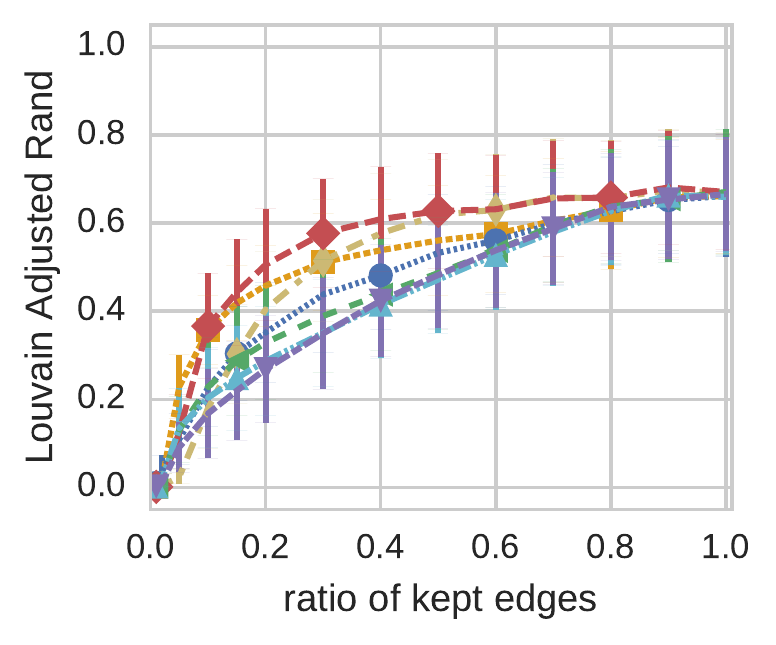}
\includegraphics[width=0.5\columnwidth]{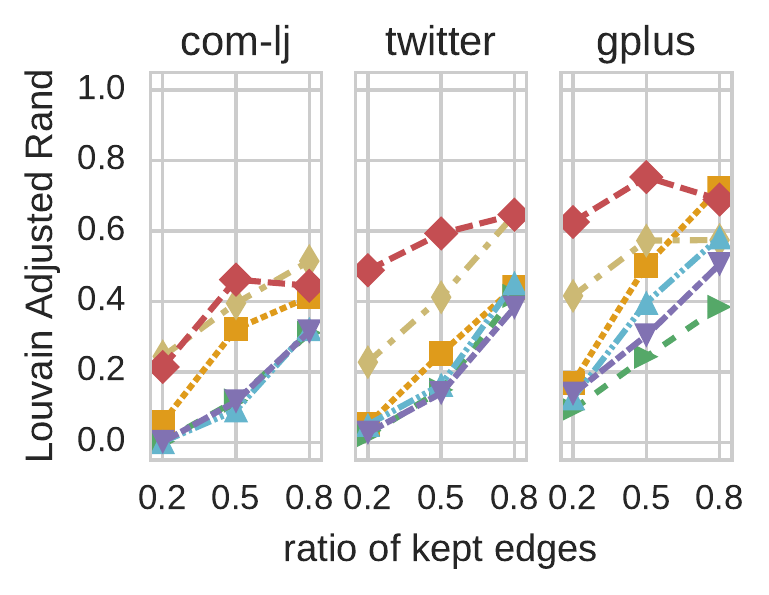}%
\includegraphics[width=0.5\columnwidth]{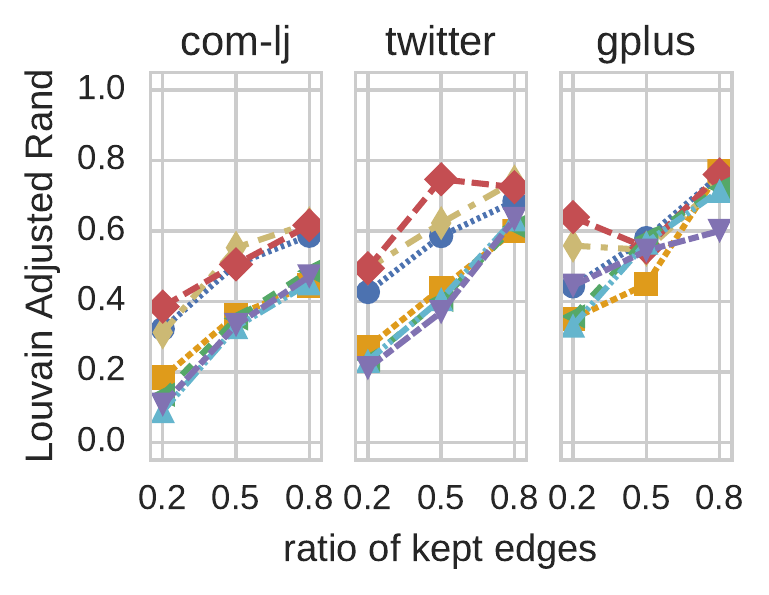}%
\end{minipage}%
\label{fig:plt_community_adjustedRandMeasure}%
}

\caption{Preservation of community structure}
\label{fig:plt_umst}

\end{center}
\end{figure}

We plot the relative inter-cluster conductance change in Figure~\ref{fig:plt_aixc_difference_percentage}.
A value of $0$ means that the conductance stays the same, a value of $-1$ indicates that the conductance became $0$ (i.e. a decrease by 100\%) and a value of $1$ indicates that the conductance has been doubled (i.e. an increase of 100\%).
We can again see that there are three categories of algorithms: the first group consisting of random edge sampling preserves the conductance values on most networks.
The second group contains only Local Degree and increases the conductance.
Edge Forest Fire has no clear behavior.
On the LiveJournal network it increases the conductance, while on Twitter and Google+ it rather decreases it.
On the Facebook networks, Edge Forest Fire without local filtering preserves the conductance values while with local filtering the conductance values are slightly increased.
The third group consisting of Jaccard Similarity, Simmelian Backbones and algebraic distance strongly decreases the conductance.
With the additional local filtering step the decrease in conductance is not as strong but still very significant.
The keeping of inter-community edges of the Local Degree method, which also explains why it preserves the connectivity so well, can be explained as follows:
Consider a hub node $x$ within a community with neighbors that are for the most part also connected to a hub node $y$ with higher degree than $x$.
Due to the way Local Degree scores edges, $x$ will lose many of its connections within the community and may be pulled into the community of a neighboring high-degree node $z$ that is not part of the original community of $x$.
Jaccard Similarity, Simmelian Backbones and algebraic distance on the other hand focus -- by design -- on intra-community edges.
Random edge sampling and Edge Forest Fire filter both types of edges almost equally distributed which is not surprising given their random nature.
Depending on the network Edge Forest Fire shows different behavior, this indicates that these networks have a different structure.

In Figure~\ref{fig:plt_average_fragmentation} it becomes obvious that only local filtering allows methods to keep the intra-cluster connectivity up to very sparse graphs.
On the Facebook networks Simmelian Backbones and Jaccard Similarity without local filtering are actually the worst in this respect, they do not keep the connectivity even though they prefer intra-cluster edges as we have seen before.
On the other networks, algebraic distance is even more extreme in this regard.
Random edge sampling and Edge Forest Fire on the non-Facebook networks are the only non-local method where a slow increase of the fragmentation can be observed, all other methods lead to a steep increase of the fragmentation during the first 10\% of edges that are removed.

Given these observations, we expect that we should still be able to find a very similar community structure at least if we use Simmelian Backbones, Jaccard Similarity or algebraic distance with local filtering.
In Figure~\ref{fig:plt_community_adjustedRandMeasure} we compare the community structure that is found by the Louvain method on the sparsified network to the one found on the original network.
For this comparison we use the adjusted rand index \cite{hubert1985comparing}.
Note that the frequently used normalized mutual information (NMI) measure reports higher similarity values for a larger number of found communities (see \eg~\cite{Vinh2009}).
This makes it unsuitable for comparing partitions on sparsified networks as we have to expect many small communities when a lot of edges are removed.
As \cite{Vinh2009} also show in their experiments, the adjusted rand index does not have these properties as it has an expected value of 0 for random partitions.

As a first observation we need to note that even when all edges are still in the network (at the right boundary of the plot for the Facebook networks), the community structure found is already different.
The Louvain method is randomized, therefore it is not unlikely that every found community structure is different.
The amount of difference between the community structures even without filtering edges indicates that there is not a single, well-defined community structure in these graphs but many different ones.
Preliminary tests with the (slightly slower) Infomap community detection algorithm \cite{rosvall2009map} which has an excellent performance on synthetic benchmark graphs for community detection \cite{lancichinetti2009community} show a very similar variance which indicates that this is not due to a weakness of the Louvain algorithm.
Filtering the edges such that we can measure that the conductance of one of these many community structures is decreased most probably does not just make this structure clearer but does also lead the algorithm into finding different community structures.
Therefore most methods lead to significantly different community structures.
It is possible that some of the sparsification methods, especially local variants of the Simmelian Backbones, Jaccard Similarity and algebraic distance, simply reveal a different community structure.
On the contrary, if all edges are kept with the same probability, the almost same set of community structures can still be found up to a certain ratio of kept edges.
Removing less than 40\% of the edges at random using random edge sampling or Edge Forest Fire does not seem to lead to more different structures than the already found ones.
Note that on the three other networks these results are hard to interpret as they are from just a single run, but the general tendencies are similar.
Local methods keep the connectivity and are thus slightly better at preserving the community structure.

\begin{figure}[tbp]

\begin{center}

\subfloat[Relative conductance change of the ground truth communities]{%
\includegraphics[width=0.5\columnwidth]{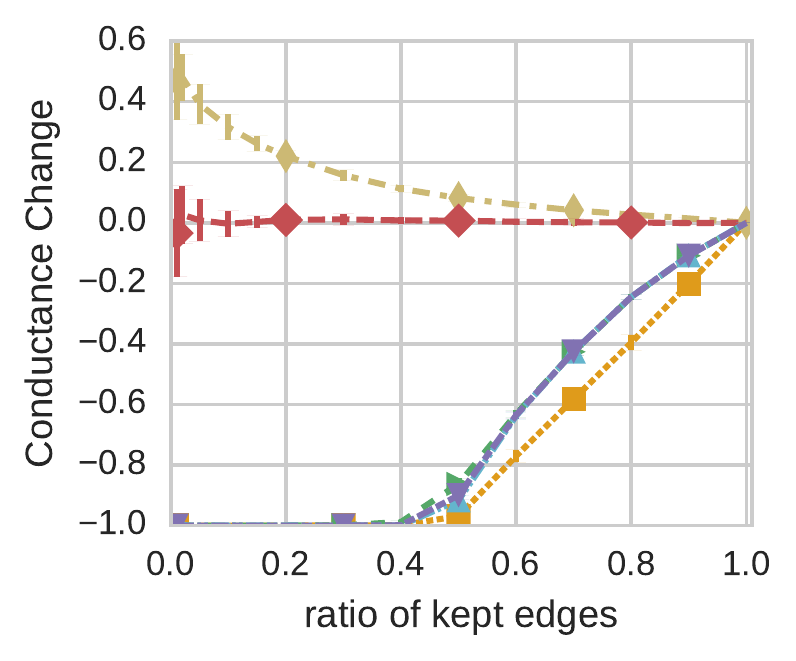}%
\includegraphics[width=0.5\columnwidth]{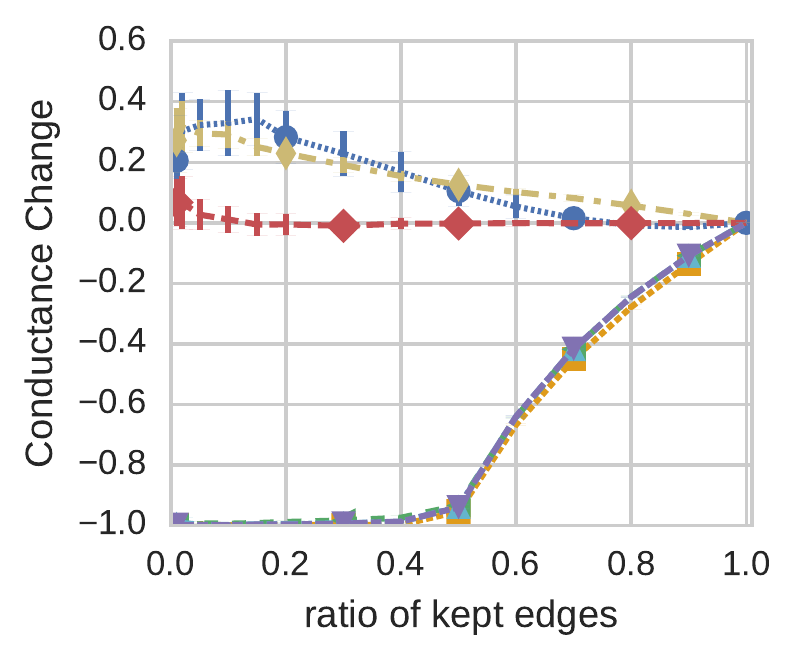}%
\label{fig:plt_aixc_percentage_lfr}%
}%

\subfloat[Average partition fragmentation of the ground truth communities]{%
\includegraphics[width=0.5\columnwidth]{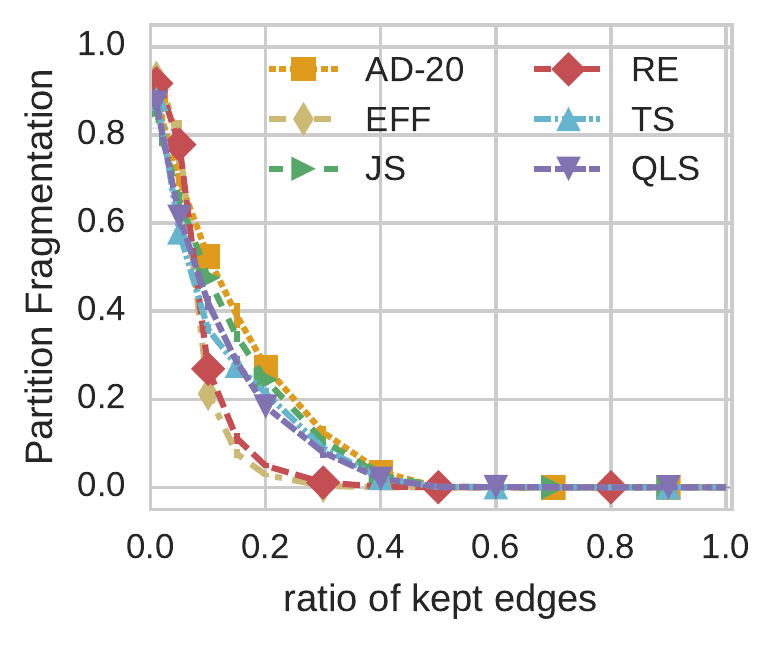}%
\includegraphics[width=0.5\columnwidth]{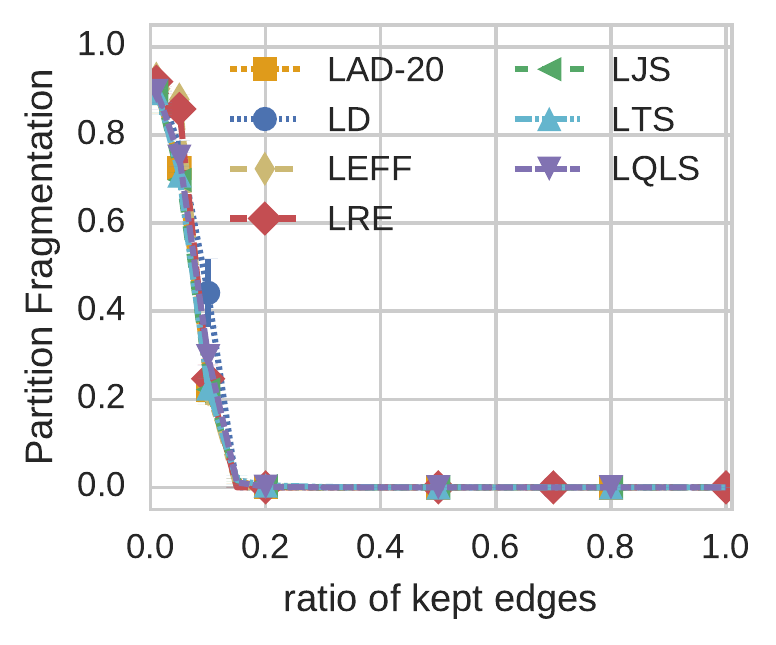}%
\label{fig:plt_average_fragmentation_lfr}%
}

\subfloat[Adjusted rand measure between ground truth communities and found communities on the sparsified network]{%
\includegraphics[width=0.5\columnwidth]{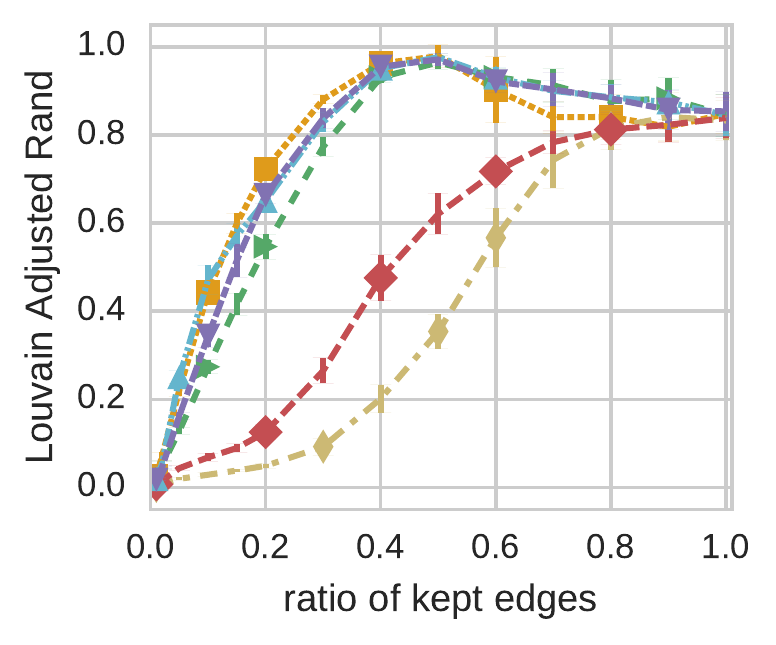}%
\includegraphics[width=0.5\columnwidth]{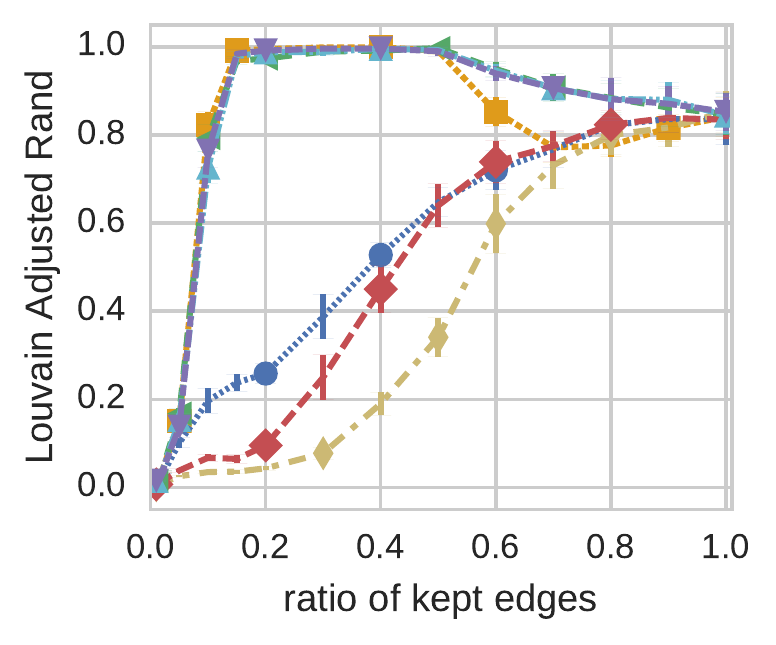}%
\label{fig:plt_adjustedRand_lfr}%
}
\caption{Preservation of the community structure on the generated LFR graphs}
\label{fig:plt_lfr}

\end{center}
\end{figure}

In order to verify the hypothesis that some differences are due to different community structures being found, we use synthetic networks with ground truth communities.
For this purpose we use the popular LFR generator~\cite{PhysRevE.80.016118}.
As parameters we choose the configuration with 1000 nodes and small communities from \cite{lancichinetti2009community}.
This means our synthetic networks have a power-law degree distribution with exponent $-2$, average degree 20 and maximum degree 50.
The communities have between 10 and 50 nodes, the community sizes also follow a power-law distribution but with exponent $-1$.
As mixing parameter $\mu$ we choose $0.5$.
This means that each node has as many neighbors in its own community as in all other communities together.
For smaller mixing parameters the differences between the different techniques are less obvious, for larger mixing parameters we reach the limits where community detection algorithms are no longer able to identify the ground truth communities.
For the plots in Figure~\ref{fig:plt_lfr} we use ten different random networks with the same configuration and report again the average and the standard deviation.
As we have known ground truth communities for these networks, we use these ground truth communities instead of a community structure found by the Louvain algorithm for the following comparisons.

For the inter-cluster conductance in Figure~\ref{fig:plt_aixc_percentage_lfr} the results are similar to the results for the Facebook networks but much clearer.
Random edge filtering almost perfectly preserves the inter-cluster conductance both in the variant without and the one with local filtering.
Edge Forest Fire and Local Degree lead to a clear increase of the conductance, again independent of the local filtering step.
Compared to the Facebook networks this increase for Edge Forest Fire is now much stronger and earlier in the sparsification process.
Simmelian Backbones, Jaccard Similarity and algebraic distance lead to a strong decrease of the inter-cluster conductance.
With 40\% remaining edges, the conductance reaches almost 0 for all of them.
Note that if a measure was able to perfectly distinguish between intra- and inter-cluster edges, the inter-cluster conductance could reach 0 when the ratio of kept edges reaches 50\%.
All methods are not far from that goal, but algebraic distance is the best method in this regard.
With a local filtering post-processing step, algebraic distance is more similar to the other methods.
For the other methods, only minor changes can be observed.
It is visible, though, that some inter-cluster edges seem to remain.

On the LFR networks the connectivity in the communities seems to be preserved much better than on the Facebook networks, see Figure~\ref{fig:plt_average_fragmentation_lfr}.
Up to 50\% of removed edges, none of the methods leads to any noticeable fragmentation.
Only when more edges are removed, the Simmelian Backbones, Jaccard Similarity and algebraic distance seem to disconnect parts of the communities.
With the additional local filtering step the connectivity inside communities is almost perfectly preserved up to 15\% remaining edges.
As the networks have an average degree of 20 we also cannot expect that connectivity is preserved much further as with 10\% remaining edges only a tree could be preserved.

In Figure~\ref{fig:plt_adjustedRand_lfr} we compare the ground truth communities to the community structure found by the Louvain algorithm (again with refinement).
Without the sparsification step, the Louvain algorithm is unable to detect the ground truth communities, this is most likely due to the resolution limit \cite{Fortunato:2007zr}.
On the generated networks with clear ground truth communities, our intuition that random edge, local degree and edge forest fire should not be able to preserve the community structure is verified.
While removing less than 20\% of the edges leads to similar detection rates, the differences increase as more and more edges are removed.
Edge Forest Fire is worst at keeping the community structure on LFR networks.
On the contrary those methods that show positive results for the preservation of the community structure actually lead to sparsified networks where the Louvain algorithm is better able to find the community structure.
As we could expect from the partition fragmentation, without local filtering the detection rate is decreased after 50\% of the edges have been removed.
With local filtering, though, there is a range between 50\% and 15\% of remaining edges where the Louvain algorithm can almost exactly recover the ground truth communities on the sparsified network.
This shows that sparsification can even increase the quality of communities found by community detection algorithms.
Algebraic distance with local filtering seems to work best in this regard.
During the first 50\% of removed edges algebraic distance shows a strange behavior, though -- especially with local filtering the detection rate first drops a bit.
This is surprising as algebraic distance leads to the strongest decrease of the inter-cluster conductance right at the beginning.
A possible explanation is that the Louvain algorithm merges especially small clusters.
If other methods filter edges between these small clusters first, this most probably helps the Louvain algorithm most.

With all these experiments we have seen that measuring the preservation of the community structure is a challenging task especially when no ground truth communities are known.
Our results suggests that the social networks either do not contain a clear community structure or that the Louvain algorithm is unable to identify this structure.
Random edge deletion and edge forest fire seem to preserve this uncertainty in the sense that the Louvain algorithm still identifies relatively similar communities.
Simmelian Backbones, Jaccard Similarity and algebraic distance on the other hand are designed to prefer intra-cluster edges which can also be seen in our experiments.
On the sparsified networks this has the effect that the Louvain algorithm detects communities that are different from the communities it detects on the original network.
On synthetic networks, local filtering with these methods preserves and even enforces the community structure. They are able to preserve the ground truth communities up to a ratio of kept edges of 0.15.
This suggests that Simmelian Backbones, Jaccard Similarity and algebraic distance with local filtering indeed keep and enforce some community structure but that on networks without clearly detectable community structure this is not necessarily the same structure as the structure that is found by the Louvain algorithm.

\subsection{Epidemic Simulations}

The previous experiments focused on static structural properties only. Now we briefly turn to dynamic, emergent properties that can be observed by simulating processes on networks.
Epidemic models are simplified means of describing the transmission of communicable diseases through a population of individuals, which can intuitively be applied to networks. Studies have recognized the importance of social networks in disease transmission~\cite{salathe2010high}. In the following we apply the SEIR model~\cite{keeling2008modeling}, which assigns one of four states to each node: Initially one node has been exposed (E) to the infection and all other nodes are susceptible (S). An exposed node becomes infectious (I) after a number of time steps. At each time step, an infectious node contacts all of its neighbors, and with a certain transmission probability, a susceptible neighbor becomes exposed. Nodes stay infectious for a given number of steps, and are then removed (R), either by immunization or death.
Counting the number of nodes of each state at each time step yields epidemic curves that describe the dynamics of the outbreak.

\begin{figure}[h]
\begin{center}
\includegraphics[width=\columnwidth]{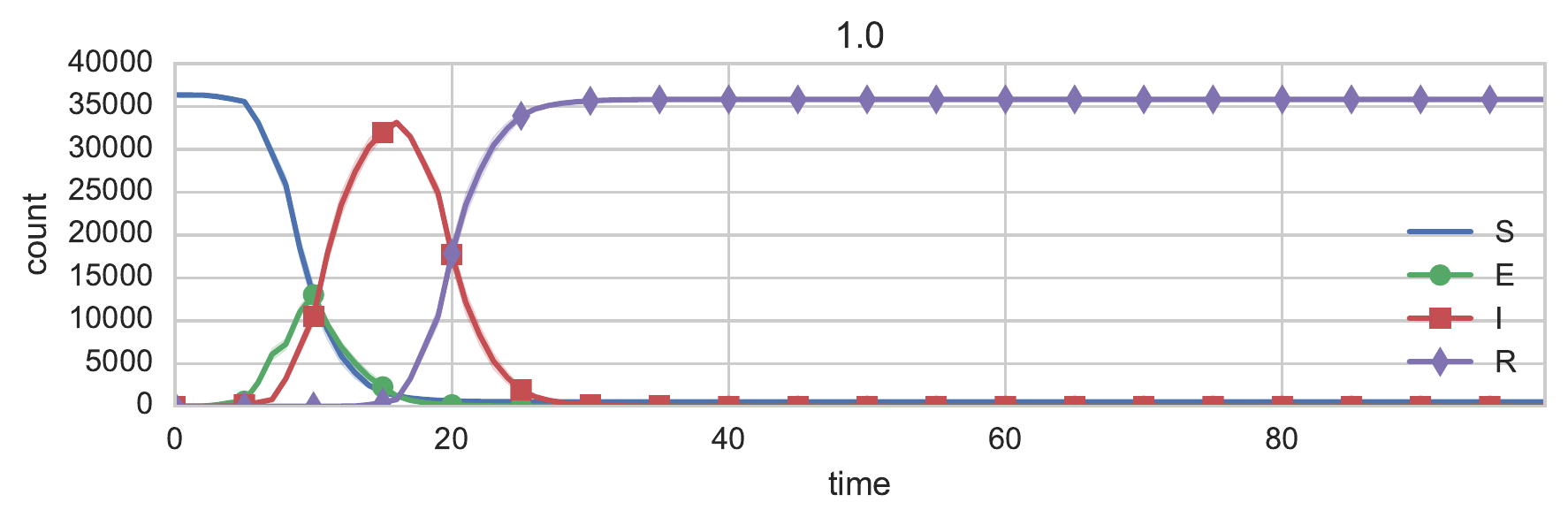}
\caption{Epidemic curves of SEIR simulation on a Facebook social network}
\label{fig:epidemic_orig}
\end{center}
\end{figure}

There is a nontrivial relationship between network structure and epidemic dynamics (\eg, they have been connected to spectral properties of the graph~\cite{wang2003epidemic}).
We can ask whether sparsified versions of a network give rise to similar epidemiological dynamics, in terms of the size and timing of a disease outbreak, and add another level of analysis for the sparsification methods.
We select \texttt{fb-Texas84} (ca. 1.6 million edges) as a representative social network and run the SEIR simulation 50 times with a latency period of  2 time steps, an infectious period of 9 time steps, and a transmission probability of 0.1. Figure~\ref{fig:epidemic_orig} shows the aggregated epidemic curves (where the central line represents the median number of nodes and the shaded areas around it the standard deviation) for the original network. While epidemic dynamics can depend strongly on the specific network structure, the following observations were roughly consistent across the Facebook-type networks.

\begin{figure}[h]
\begin{center}
\includegraphics[width=\columnwidth]{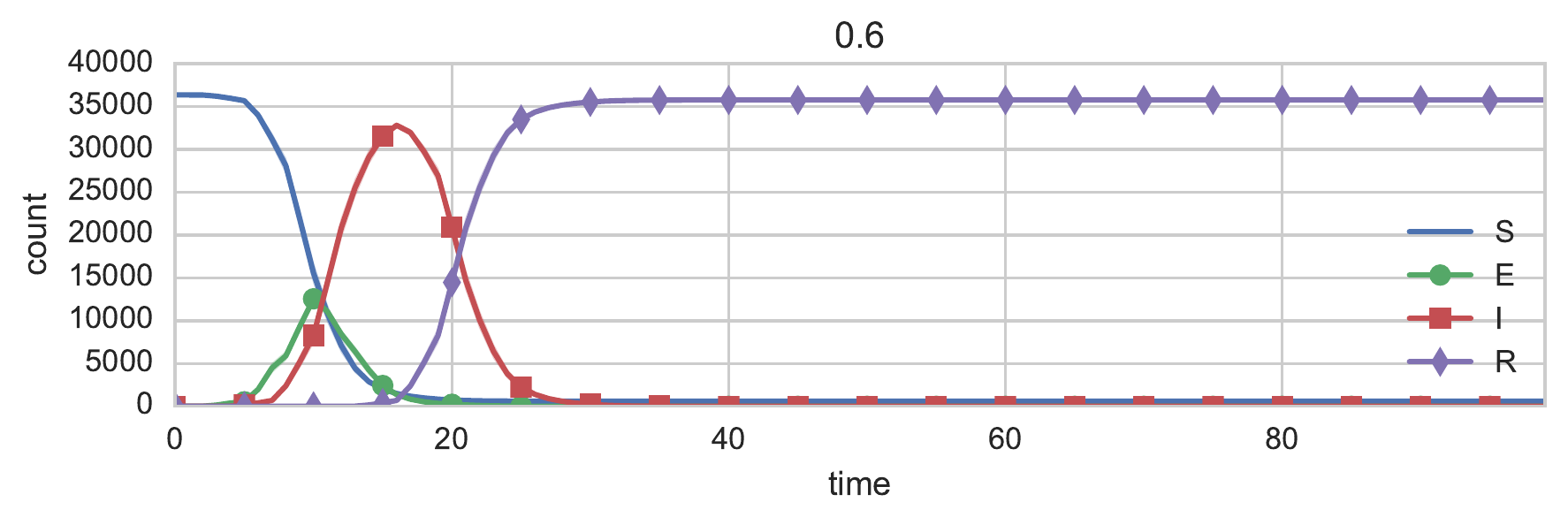}
\includegraphics[width=\columnwidth]{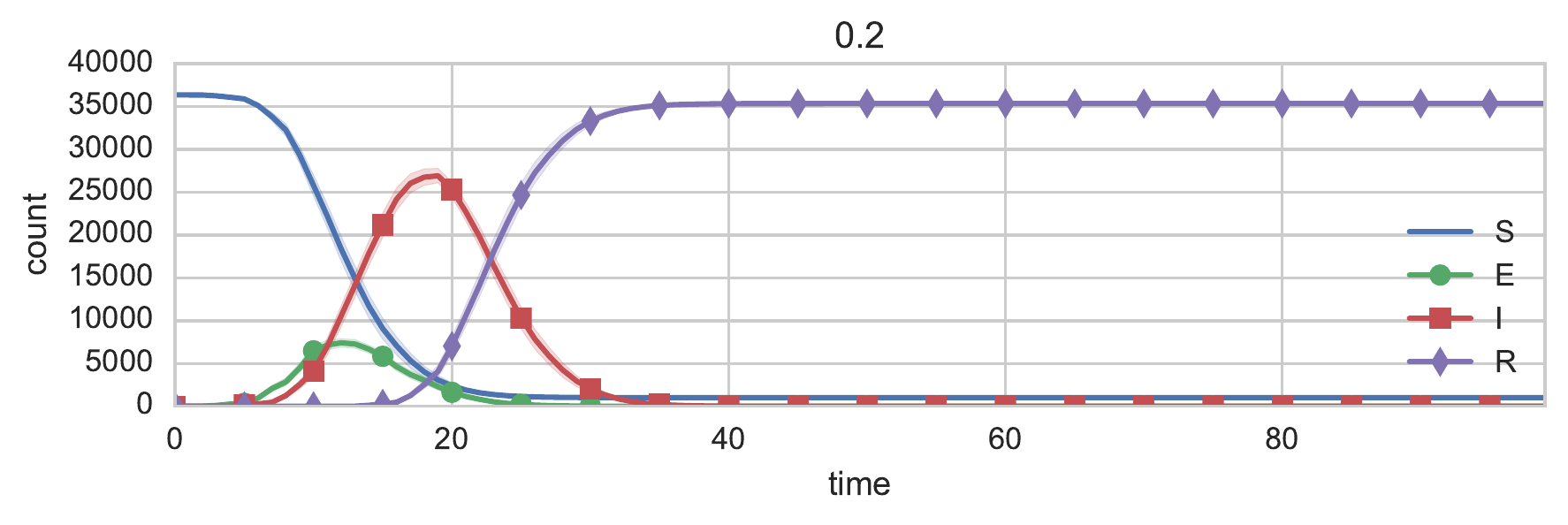}
\caption{Epidemic curves after Local Degree sparsification}
\label{fig:epidemic_LD}
\end{center}
\end{figure}

\begin{figure}[h]
\begin{center}
\includegraphics[width=\columnwidth]{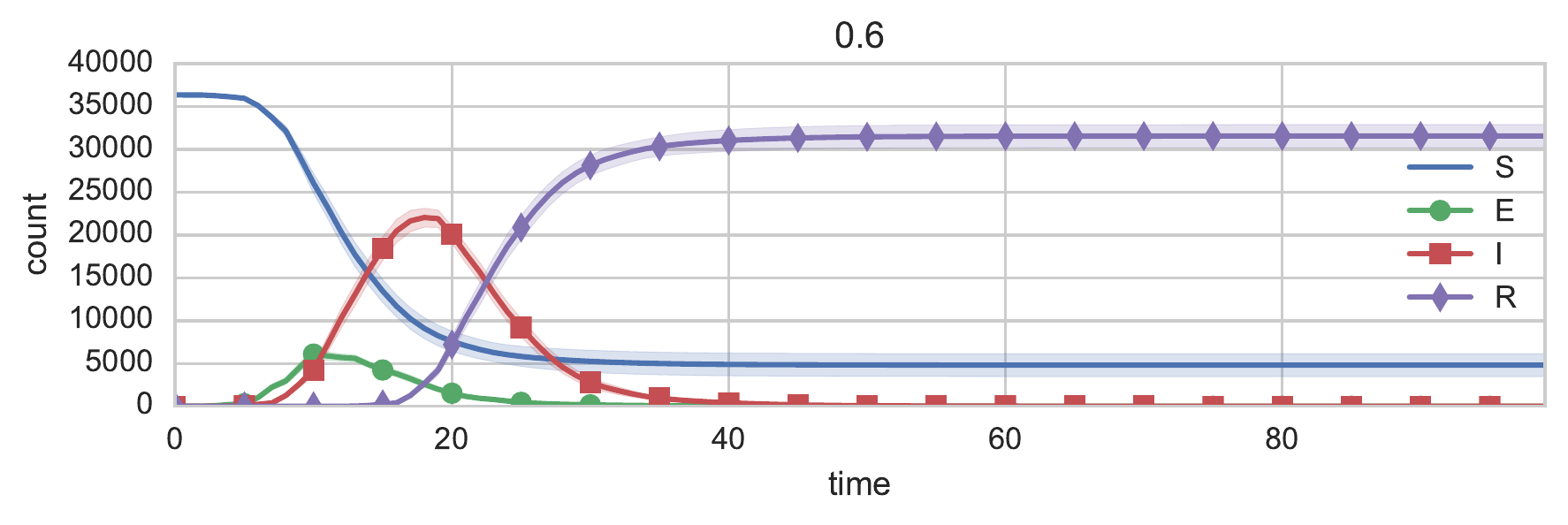}
\includegraphics[width=\columnwidth]{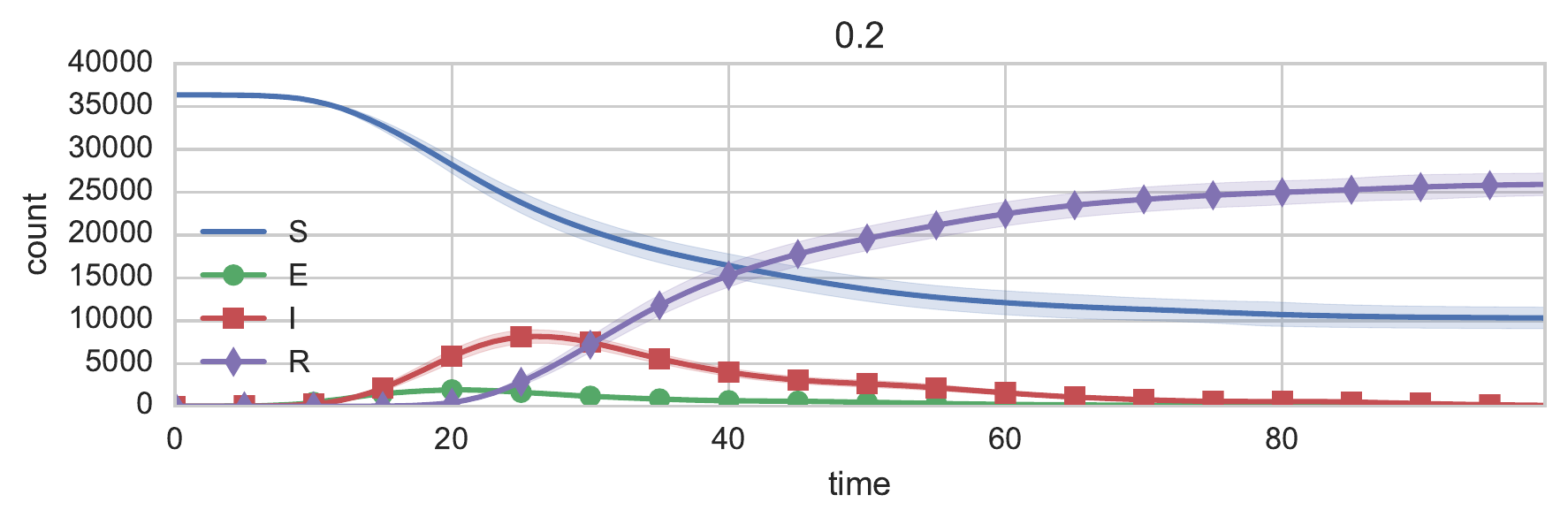}
\caption{Epidemic curves after algebraic distance sparsification}
\label{fig:epidemic_AD}
\end{center}
\end{figure}

The Local Degree method most closely replicates the epidemic curves of the original down to an edge ratio of 0.2, producing only a minor delay in the outbreak and slightly lower peak number of infected nodes, but an identical converged state (Fig.~\ref{fig:epidemic_LD}).
One reason for this is certainly that connectedness and short paths are preserved. It may also point to the importance of local hubs in epidemic propagation.
Random edge sampling and Forest Fire sparsification also perform well and produce a similar high fidelity.
Other methods deviate more from the original epidemic dynamics by delaying and dampening the outbreak, with some also strongly reducing the final number of infections. 
For Local Jaccard Similarity, thinning the network slows the outbreak slightly, leading to a less sharp and high peak of infected nodes,  but reproduces essentially the same epidemic curve shapes and final state.
At the other end of the spectrum, sparsification by algebraic distance (Fig.~\ref{fig:epidemic_AD}) and the Simmelian methods selectively removes bottleneck edges between dense regions of the network. These edges are likely to be critically important for the propagation of a disease, and hence epidemic dynamics are significantly altered with deleting those edges.

\subsection{Running Time}

\begin{figure}[htbp]
\begin{center}
\includegraphics[width=.85\columnwidth]{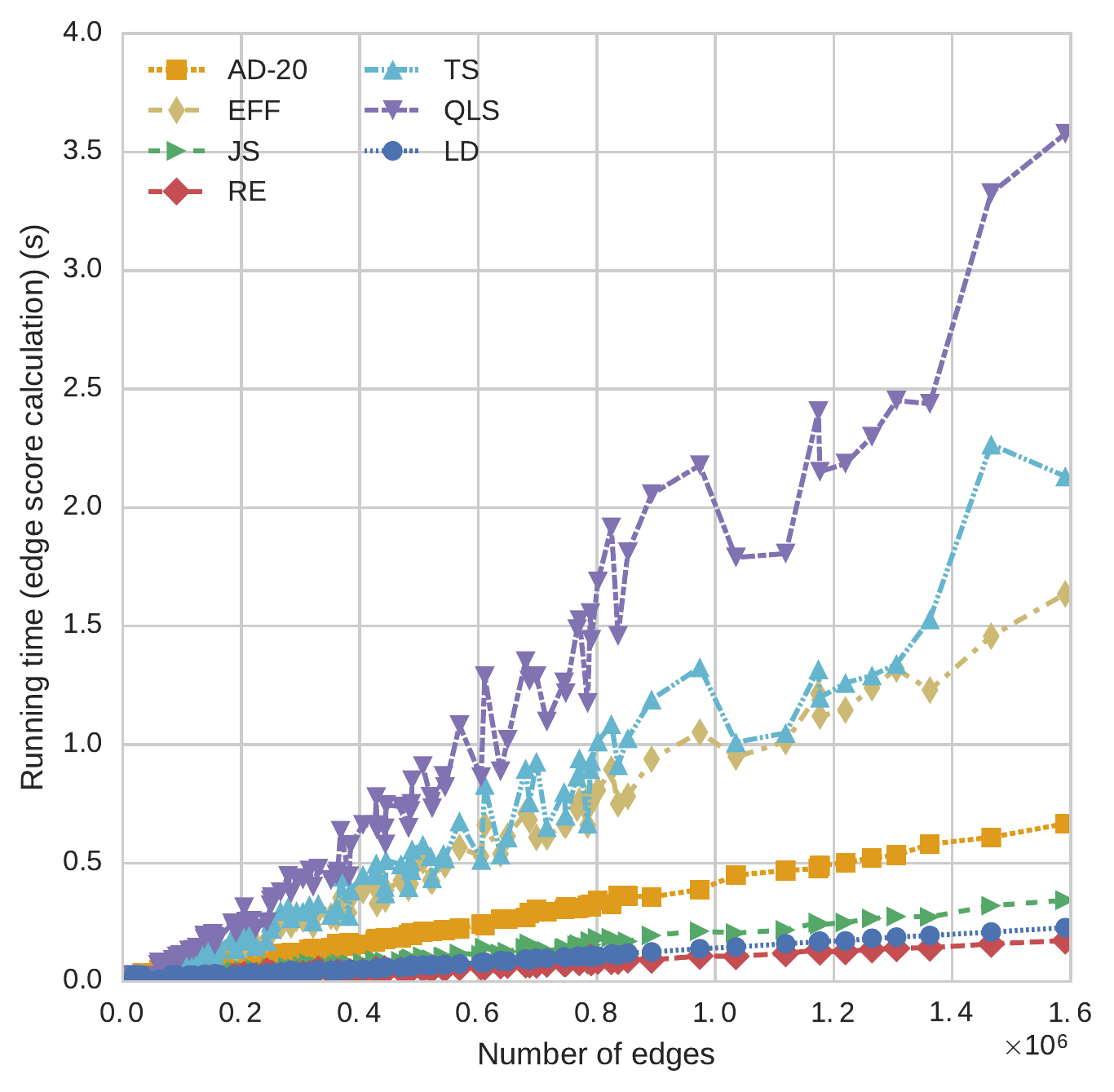}
\caption{Running times of various edge scoring methods on the Facebook networks}
\label{fig:plt_runtimes}
\end{center}
\end{figure}

Measured running times are shown in Fig.~\ref{fig:plt_runtimes}.
Random Edge sparsification is clearly the fastest method, closely followed by Local Degree.
Jaccard Similarity is also not much slower and scales also very well.
Therefore these methods are well suited for large-scale networks in the range of millions to billions of edges.
The efficiency of the Jaccard Similarity method shows that our parallel triangle counting implementation is indeed very scalable.
The authors also proposed inexact Jaccard coefficient calculation for a further speedup though given our numbers it can be doubted if -- given an efficient triangle counter -- this is necessary or helpful at all.
Algebraic distance is a bit slower but scales very well nevertheless.
Using less systems or iterations could further speed-up algebraic distance if speed is an issue.
Both Simmelian methods are significantly slower than the other methods, but still efficient enough for the network sizes we consider.
The visible difference between quadrilateral and triangular Simmelian Backbones can be explained by the difference between triangle and quadrangle counting, additionally we did not parallelize the latter.
While the time complexity in O-notation of Edge Forest Fire is difficult to assess, it seems to be slightly faster than Simmelian Backbones.

\section{Conclusion}

Our experimental study on networks from Facebook, Twitter and Google+ as well as synthentically generated networks shows that several sparsification methods are capable of preserving a set of relevant properties of social networks when up to 80\% of edges have been removed.

Random edge deletion performs surprisingly well and retains a wide range of properties, but more targeted methods can perform even better.
We propose local filtering as a generally applicable and computationally cheap post-processing step for edge sparsification methods that improves the preservation of almost all properties as it leads to a more equal rate of filtering across the network.
Simmelian Backbones, Jaccard Similarity and algebraic distance prefer intra-cluster edges and thus do not keep global structures but with the added local filtering step they are able to enforce and retain a community structure as it was already shown for Jaccard Similarity. However, the preserved community structure is not necessarily the same as the one the Louvain algorithm finds.
Our novel method Local Degree, which is based on the notion that connections to hubs are highly important for the network's structure, in contrast preserves shortest paths and the overall connectivity of the network.
This can be seen at the almost perfectly preserved diameter and the well-preserved behavior of the network in epidemic simulations.
Depending on the network, the Local Degree method is also able to preserve clustering coefficients and centralities.
Our adaption of the Forest Fire sampling algorithm to edge scoring depends strongly on the specific network's structure.
It is good at preserving connectivity, on some networks it also preserves centralities and the diameter.

We hope that the conceptual framework of edge scoring and filtering as well as our evaluation methods are steps towards a more unified perspective on a variety of related methods that have been proposed in different contexts.
Future developments can be easily carried out within this framework and based on our implementations, which are available as part of the open-source network analysis package NetworKit\footnotemark.
\footnotetext{\url{https://networkit.iti.kit.edu/}}

\subsubsection*{{\scriptsize{Acknowledgements.}}}

\begin{scriptsize}
This work is partially supported by the German Research Foundation (DFG) under grants ME~3619/3-1 and WA~654/22-1 within the Priority Programme 1736 \emph{Algorithms for Big Data}.
\end{scriptsize}

\bibliographystyle{apalike}
\bibliography{Bibliography}

\begin{thebibliography}{}

\bibitem[Ahmed et~al., 2014]{ahmed2014network}
Ahmed, N.~K., Neville, J., and Kompella, R. (2014).
\newblock Network sampling: From static to streaming graphs.
\newblock {\em ACM Transactions on Knowledge Discovery from Data (TKDD)},
  8(2):7.

\bibitem[Barab{\'a}si and Albert, 1999]{ba-esrn-99}
Barab{\'a}si, A.-L. and Albert, R. (1999).
\newblock {Emergence of scaling in random networks}.
\newblock {\em Science}, 286:509--512.

\bibitem[Bastian et~al., 2009]{BastianHJ09Gephi}
Bastian, M., Heymann, S., and Jacomy, M. (2009).
\newblock Gephi: An open source software for exploring and manipulating
  networks.
\newblock In Adar, E., Hurst, M., Finin, T., Glance, N.~S., Nicolov, N., and
  Tseng, B.~L., editors, {\em ICWSM}. The AAAI Press.

\bibitem[Batson et~al., 2013]{batson2013spectral}
Batson, J., Spielman, D.~A., Srivastava, N., and Teng, S.-H. (2013).
\newblock Spectral sparsification of graphs: theory and algorithms.
\newblock {\em communications of the ACM}, 56(8):87--94.

\bibitem[Borassi et~al., 2015]{Borassi2015}
Borassi, M., Crescenzi, P., Habib, M., Kosters, W.~A., Marino, A., and Takes,
  F.~W. (2015).
\newblock Fast diameter and radius bfs-based computation in (weakly connected)
  real-world graphs: With an application to the six degrees of separation
  games.
\newblock {\em Theoretical Computer Science}, 586:59 -- 80.
\newblock Fun with Algorithms.

\bibitem[Chen and Safro, 2011]{chen2011algebraic}
Chen, J. and Safro, I. (2011).
\newblock Algebraic distance on graphs.
\newblock {\em SIAM Journal on Scientific Computing}, 33(6):3468--3490.

\bibitem[Chiba and Nishizeki, 1985]{Chi85}
Chiba, N. and Nishizeki, T. (1985).
\newblock Arboricity and subgraph listing algorithms.
\newblock {\em SIAM Journal on Computing}, 14(1):210--223.

\bibitem[Costa et~al., 2011]{costa2011analyzing}
Costa, L. d.~F., Oliveira~Jr, O.~N., Travieso, G., Rodrigues, F.~A.,
  Villas~Boas, P.~R., Antiqueira, L., Viana, M.~P., and Correa~Rocha, L.~E.
  (2011).
\newblock Analyzing and modeling real-world phenomena with complex networks: a
  survey of applications.
\newblock {\em Advances in Physics}, 60(3):329--412.

\bibitem[Ebbes et~al., 2008]{ebbes2008sampling}
Ebbes, P., Huang, Z., Rangaswamy, A., Thadakamalla, H.~P., and Unit, O. R.
  G.~B. (2008).
\newblock Sampling large-scale social networks: Insights from simulated
  networks.
\newblock In {\em 18th Annual Workshop on Information Technologies and Systems,
  Paris, France}. Citeseer.

\bibitem[Fortunato and Barthelemy, 2007]{Fortunato:2007zr}
Fortunato, S. and Barthelemy, M. (2007).
\newblock Resolution limit in community detection.
\newblock {\em Proceedings of the National Academy of Sciences}, 104(1):36--41.

\bibitem[Fortunato et~al., 2008]{fortunato2008approximating}
Fortunato, S., Bogu{\~n}{\'a}, M., Flammini, A., and Menczer, F. (2008).
\newblock Approximating pagerank from in-degree.
\newblock In {\em Algorithms and Models for the Web-Graph}, pages 59--71.
  Springer.

\bibitem[Geisberger et~al., 2008]{geisberger2008better}
Geisberger, R., Sanders, P., and Schultes, D. (2008).
\newblock Better approximation of betweenness centrality.
\newblock In {\em ALENEX}, pages 90--100. SIAM.

\bibitem[Gleiser and Danon, 2003]{GleiserJazz}
Gleiser, P.~M. and Danon, L. (2003).
\newblock Community structure in jazz.
\newblock {\em Advances in Complex Systems}, 6(4):565--574.

\bibitem[Hubert and Arabie, 1985]{hubert1985comparing}
Hubert, L. and Arabie, P. (1985).
\newblock Comparing partitions.
\newblock {\em Journal of Classification}, 2(1):193--218.

\bibitem[Keeling and Rohani, 2008]{keeling2008modeling}
Keeling, M. and Rohani, P. (2008).
\newblock {\em Modeling infectious diseases in humans and animals}.
\newblock Princeton University Press.

\bibitem[Lancichinetti and Fortunato, 2009a]{PhysRevE.80.016118}
Lancichinetti, A. and Fortunato, S. (2009a).
\newblock Benchmarks for testing community detection algorithms on directed and
  weighted graphs with overlapping communities.
\newblock {\em Phys. Rev. E}, 80(1):016118.

\bibitem[Lancichinetti and Fortunato, 2009b]{lancichinetti2009community}
Lancichinetti, A. and Fortunato, S. (2009b).
\newblock Community detection algorithms: A comparative analysis.
\newblock {\em Phys. Rev. E}, 80:056117.

\bibitem[Leskovec and Faloutsos, 2006]{Leskovec2006}
Leskovec, J. and Faloutsos, C. (2006).
\newblock Sampling from large graphs.
\newblock In {\em Proceedings of the 12th ACM SIGKDD International Conference
  on Knowledge Discovery and Data Mining}, KDD '06, pages 631--636, New York,
  NY, USA. ACM.

\bibitem[Leskovec and Mcauley, 2012]{leskovec2012circles}
Leskovec, J. and Mcauley, J.~J. (2012).
\newblock Learning to discover social circles in ego networks.
\newblock In Pereira, F., Burges, C., Bottou, L., and Weinberger, K., editors,
  {\em Advances in Neural Information Processing Systems 25}, pages 539--547.
  Curran Associates, Inc.

\bibitem[Lindner et~al., 2015]{DBLP:conf/asunam/LindnerSHMW15}
Lindner, G., Staudt, C.~L., Hamann, M., Meyerhenke, H., and Wagner, D. (2015).
\newblock Structure-preserving sparsification of social networks.
\newblock In Pei, J., Silvestri, F., and Tang, J., editors, {\em Proceedings of
  the 2015 {IEEE/ACM} International Conference on Advances in Social Networks
  Analysis and Mining, {ASONAM} 2015, Paris, France, August 25 - 28, 2015},
  pages 448--454. {ACM}.

\bibitem[Newman, 2010]{newman2010networks}
Newman, M. (2010).
\newblock {\em Networks: an introduction}.
\newblock Oxford University Press.

\bibitem[Nick et~al., 2013]{Nick13}
Nick, B., Lee, C., Cunningham, P., and Brandes, U. (2013).
\newblock Simmelian backbones: Amplifying hidden homophily in facebook
  networks.
\newblock In {\em Proceedings of the 2013 IEEE/ACM International Conference on
  Advances in Social Networks Analysis and Mining}, ASONAM '13, pages 525--532,
  New York, NY, USA. ACM.

\bibitem[Nocaj et~al., 2014]{nocaj2014hairballs}
Nocaj, A., Ortmann, M., and Brandes, U. (2014).
\newblock Untangling hairballs - from 3 to 14 degrees of separation.
\newblock In Duncan, C.~A. and Symvonis, A., editors, {\em Graph Drawing - 22nd
  International Symposium, {GD} 2014, W{\"{u}}rzburg, Germany, September 24-26,
  2014, Revised Selected Papers}, volume 8871 of {\em Lecture Notes in Computer
  Science}, pages 101--112. Springer.

\bibitem[Ortmann and Brandes, 2014]{ortmann2014triangle}
Ortmann, M. and Brandes, U. (2014).
\newblock Triangle listing algorithms: Back from the diversion.
\newblock In McGeoch, C.~C. and Meyer, U., editors, {\em 2014 Proceedings of
  the Sixteenth Workshop on Algorithm Engineering and Experiments, {ALENEX}
  2014, Portland, Oregon, USA, January 5, 2014}, pages 1--8. {SIAM}.

\bibitem[Page et~al., 1999]{Page1999}
Page, L., Brin, S., Motwani, R., and Winograd, T. (1999).
\newblock The pagerank citation ranking: Bringing order to the web.
\newblock Technical Report 1999-66, Stanford InfoLab.
\newblock Previous number = SIDL-WP-1999-0120.

\bibitem[Rosvall et~al., 2009]{rosvall2009map}
Rosvall, M., Axelsson, D., and Bergstrom, C.~T. (2009).
\newblock The map equation.
\newblock {\em The European Physical Journal Special Topics}, 178(1):13--23.

\bibitem[Saha et~al., 2013]{saha2013sparsification}
Saha, T., Rangwala, H., and Domeniconi, C. (2013).
\newblock Sparsification and sampling of networks for collective
  classification.
\newblock In {\em Social Computing, Behavioral-Cultural Modeling and
  Prediction}, pages 293--302. Springer.

\bibitem[Salath{\'e} et~al., 2010]{salathe2010high}
Salath{\'e}, M., Kazandjieva, M., Lee, J.~W., Levis, P., Feldman, M.~W., and
  Jones, J.~H. (2010).
\newblock A high-resolution human contact network for infectious disease
  transmission.
\newblock {\em Proceedings of the National Academy of Sciences},
  107(51):22020--22025.

\bibitem[Satuluri et~al., 2011]{Satuluri2011}
Satuluri, V., Parthasarathy, S., and Ruan, Y. (2011).
\newblock Local graph sparsification for scalable clustering.
\newblock In {\em Proceedings of the 2011 ACM SIGMOD International Conference
  on Management of Data}, SIGMOD '11, pages 721--732, New York, NY, USA. ACM.

\bibitem[Serrano et~al., 2009]{Serrano09}
Serrano, M.~{\'A}., Bogu{\~n}{\'a}, M., and Vespignani, A. (2009).
\newblock Extracting the multiscale backbone of complex weighted networks.
\newblock {\em Proceedings of the National Academy of Sciences},
  106(16):6483--6488.

\bibitem[Shun and Tangwongsan, 2015]{shun2015multicore}
Shun, J. and Tangwongsan, K. (2015).
\newblock Multicore triangle computations without tuning.
\newblock In {\em Proceedings of the IEEE International Conference on Data
  Engineering (ICDE)}.

\bibitem[Simmel and Wolff, 1950]{simmel1950sociology}
Simmel, G. and Wolff, K. (1950).
\newblock {\em The Sociology of Georg Simmel}.
\newblock Free Press paperback. Free Press.

\bibitem[Staudt and Meyerhenke, 2015]{staudt2015engineering}
Staudt, C. and Meyerhenke, H. (2015).
\newblock Engineering parallel algorithms for community detection in massive
  networks.
\newblock {\em Parallel and Distributed Systems, IEEE Transactions on},
  PP(99):1--1.

\bibitem[Staudt et~al., 2014]{DBLP:journals/corr/StaudtSM14}
Staudt, C., Sazonovs, A., and Meyerhenke, H. (2014).
\newblock Networkit: A tool suite for large-scale complex network analysis.
\newblock {\em CoRR}, abs/1403.3005.

\bibitem[Traud et~al., 2012]{traud2012social}
Traud, A.~L., Mucha, P.~J., and Porter, M.~A. (2012).
\newblock Social structure of facebook networks.
\newblock {\em Physica A: Statistical Mechanics and its Applications},
  391(16):4165--4180.

\bibitem[Vinh et~al., 2009]{Vinh2009}
Vinh, N.~X., Epps, J., and Bailey, J. (2009).
\newblock Information theoretic measures for clusterings comparison: Is a
  correction for chance necessary?
\newblock In {\em Proceedings of the 26th Annual International Conference on
  Machine Learning}, ICML '09, pages 1073--1080, New York, NY, USA. ACM.

\bibitem[Wang et~al., 2003]{wang2003epidemic}
Wang, Y., Chakrabarti, D., Wang, C., and Faloutsos, C. (2003).
\newblock Epidemic spreading in real networks: An eigenvalue viewpoint.
\newblock In {\em Reliable Distributed Systems, 2003. Proceedings. 22nd
  International Symposium on}, pages 25--34. IEEE.

\bibitem[Yang and Leskovec, 2012]{Yang:2012fk}
Yang, J. and Leskovec, J. (2012).
\newblock Defining and evaluating network communities based on ground-truth.
\newblock In {\em Proceedings of the ACM SIGKDD Workshop on Mining Data
  Semantics}, page~3. ACM.

\end{thebibliography}

\end{document}